\documentclass[aps,twocolumn,prl]{revtex4-1}

\usepackage{graphicx}
\usepackage{upgreek}
\usepackage{amsmath}
\usepackage{amssymb}
\usepackage{chngpage}
\usepackage{tabularx}
\usepackage{array,booktabs}
\newcolumntype{C}{>{\centering\arraybackslash}X}
\usepackage{nicefrac}
\usepackage{color,soul}
\usepackage{xcolor}
\usepackage{enumitem}

\begin{document}
\title{Molecular Dynamics Simulations of NMR Relaxation and Diffusion of Heptane Confined in a Polymer Matrix}
\author{Arjun Valiya Parambathu}
\author{Philip M. Singer}
\author{George J. Hirasaki}
\author{Walter G. Chapman}
\author{Dilip Asthagiri}
\email{dilip.asthagiri@rice.edu}
\affiliation{Department of Chemical and Biomolecular Engineering, Rice University, 6100 Main St., Houston, Texas 77005, USA}
\date{\today}


\begin{abstract}
The mechanism behind the NMR surface relaxation and the large $T_1$/$T_2$ ratio of light hydrocarbons confined in the nano-pores of kerogen remains poorly understood, and consequently has engendered much debate. Towards bringing a molecular-scale resolution to this problem, we present molecular dynamics (MD) simulations of $^1$H NMR relaxation and diffusion of heptane in a polymer matrix, where the high-viscosity polymer is a model for kerogen and bitumen that provides an organic ``surface" for heptane. 
We calculate the autocorrelation function $G(t)$ for $^1$H-$^1$H dipole-dipole interactions of heptane in the polymer matrix and use this
to generate the NMR frequency ($f_0$) dependence of $T_1$ and $T_2$ relaxation times as a function of $\phi_{C7}$.  We find that increasing molecular confinement increases the correlation time of the heptane molecule, which decreases the surface relaxation times for heptane in the polymer matrix. For weak confinement ($\phi_{C7} > 50$ vol\%), we find that $T_{1S}/T_{2S} \simeq 1$. 
Under strong confinement ($\phi_{C7} \lesssim $ 50 vol\%), we find that the ratio $T_{1S}/T_{2S} \gtrsim 4$ increases with decreasing $\phi_{C7}$, and that the dispersion relation $T_{1S} \propto f_0$ is consistent with previously reported measurements of polymers and bitumen. Such frequency dependence in bitumen has been previously attributed to paramagnetism, but our
studies suggests that $^1$H-$^1$H dipole-dipole interactions enhanced by organic nano-pore confinement dominates the NMR response in saturated organic-rich shales, without the need to invoke paramagnetism.
\end{abstract}

\maketitle

\section{Introduction}

The traditional interpretation of $^1$H NMR relaxation times $T_1$ and $T_2$ (and implicitly the autocorrelation function $G(t)$ for $^1$H-$^1$H dipole-dipole interactions) rely on the Bloembergen, Purcell, and Pound (BPP) \cite{bloembergen:pr1948} theory for intra-molecular relaxation by rotational diffusion, Torrey \cite{torrey:pr1953} and Hwang and Freed \cite{hwang:JCP1975} for inter-molecular relaxation by translational diffusion, and, Hubbard \cite{hubbard:pr1963} and Bloom \cite{bloom:cjp1967b} for spin-rotation relaxation. These early pioneering theories were nevertheless built on strong assumptions regarding the molecular structure and interaction, such as the assumption of rigid molecules without internal motions. In this regard, theories for the effects of internal motions on $T_1$ and $T_2$ of ellipsoidal molecules were developed by Woessner \cite{woessner:jcp1965}, and later a more general phenomenological model was developed by Lipari and Szabo \cite{lipari:jacs1982,lipari:jacs1982b}. 

Until recently, petro-physicists have been forced to rely on these classical theories to interpret $T_1$ and $T_2$ of complex fluids such as crude-oils and bitumen. One of the biggest conundrums has been that at high viscosity, the log-mean $T_{1LM}$ becomes independent of viscosity over temperature ($\eta/T$), and (roughly) proportional to the NMR frequency $f_0$, i.e. $T_{1LM} \propto f_0$, for heavy crude-oils and bitumen \cite{vinegar:spefe1991,latorraca:spwla1999,zhang:spwla2002,yang:jmr2008,yang:petro2012,kausik:petro2019}. To address this, a new phenomenological model \cite{singer:SPWLA2017,singer:EF2018} was developed which uses the Lipari and Szabo model, plus it takes into account the multi-component nature of complex reservoir fluids by modifying the exponent of the frequency dependence in the BPP model. This new phenomenological model successfully accounted for the viscosity and frequency dependence of $T_{1LM}$ for heavy crude-oils, bitumen, and polymer-heptane mixes, without invoking paramagnetism.

Another conundrum has been that at high viscosity, the log-mean $T_{2LM}$ of heavy crude-oils and bitumen has a viscosity dependence of $T_{2LM} \propto (\eta/T)^{-0.5}$ \cite{yang:jmr2008,yang:petro2012,kausik:petro2019}, which was also observed for pure polymers \cite{singer:EF2018}. Several models have been proposed to account for this behavior \cite{korb:jpcc2015,kausik:petro2019}, yet no consensus has yet been reached. Overall, it is clear that the viscosity and frequency dependence of $T_{1LM}$ and $T_{2LM}$ present significant deviations from the classical theories of NMR relaxation of bulk fluids, and that further investigations are required. 

A better theoretical understanding of the observed relaxation $T_{1,2}$ in fluids can be achieved if one can split the different relaxation mechanisms such as intra-molecular, inter-molecular, and spin-rotation relaxation. One way to do this experimentally is to perform $^1$H or $^2$H NMR on partially deuterated molecules (i.e. partially replace $^1$H with $^2$H), such as in the case of glycerol \cite{meier:jcp2014}, polymer melts \cite{hofmann:macro2014}, or methane \cite{bloom:cjp1967b}. This technique assumes that deuterating does not alter the local molecular dynamics due to changes in the rotational and vibrational modes of the molecule. In the case of methane, the symmetry of the molecule is altered by deuteration, which complicates the interpretation. 

Molecular dynamics (MD) simulations have already played a helpful, guiding role in this regard. We have already shown that MD simulations of $^1$H-$^1$H dipole-dipole relaxation can naturally separate intra-molecular from inter-molecular $T_{1,2}$ for liquid-state $n$-alkanes and water \cite{singer:jmr2017,singer:jcp2018,asthagiri:seg2018}, as well as $^1$H spin-rotation relaxation for methane \cite{singer:jcp2018b}. Our MD simulations have also shown the effects of internal motions and molecular geometry on $T_{1,2}$ of various hydrocarbons, as well as the differences in $T_{1,2}$ between methyl and methylene $^1$H's across the $n$-alkane chain \cite{singer:jcp2018,asthagiri:seg2018}. MD simulations have shed light on the limitations of the classical theories, and they significantly advanced our understanding of $T_{1,2}$ of bulk fluids, without any free parameters or models in the interpretation of the MD simulations.

Aside from bulk fluids, another major conundrum in petro-physics is the origin of the NMR surface relaxation $T_{1S}$ and $T_{2S}$ of fluids confined in organic-rich shales \cite{ozen:petro2013,rylander:spe2013,jiang:spwla2013,singer:sca2013,washburn:jmr2013,kausik:sca2014,daigle:urtec2014,washburn:cmr2014,korb:jpcc2014,nicot:petro2015,lessenger:SPWLA2015,birdwell:ef2015,fleury:jpse2016,kausik:petro2016,singer:petro2016,sun:spwla2016,sondergeld:spwla2016,yang:ef2016,zhang:geo2017,romero:ijcg2017,chen:petro2017,valori:ef2017,washburn:jmr2017,thern:mmm2018,korb:mmm2018,tandon:spwla2019,xie:spwla2019,song:pnmrs2019,faux:mol2019,ye:spe2019,dang:spe2019,kanwar:urtec2019,zhu:EF2019,wang:urtec2019}. For instance, Ozen and Sigal \cite{ozen:petro2013} first reported that light hydrocarbons exhibit large ratios $T_{1S}/T_{2S} \gtrsim 4$ when confined in organic-shale, while small ratios $T_{1S}/T_{2S} \simeq 2$ are found for water. While this provides a good contrast mechanism for fluid typing and saturation estimates in organic-rich shale, the mechanism behind this observation is still not well understood. In some cases, $^1$H-$^1$H dipole-dipole relaxation is thought to be the dominant mechanism \cite{washburn:cmr2014,singer:petro2016,fleury:jpse2016,zhang:geo2017,washburn:jmr2017,tandon:spwla2019,xie:spwla2019}, while in other cases surface-paramagnetism is thought to be the dominant mechanism \cite{korb:jpcc2014,nicot:petro2015,korb:mmm2018,zhu:EF2019}. 

In order to shed light on this subject, we present MD simulations of $T_{1}$ and $T_{2}$ of heptane in polymer-heptane mixtures, where the high-viscosity polymer is a model for kerogen and bitumen that provides an organic ``surface" and organic transient ``pores" for heptane. Our premise behind this analogy is that the organic matter in kerogen and bitumen is essentially made up of cross-linked polymers (plus aromatics), where cross-linking turns the liquid polymers into highly-viscous bitumen or solid kerogen when more cross-links are present. We simulate the surface-relaxivity parameters $\rho_1$ and $\rho_2$ for heptane, and the surface-relaxation ratio $T_{1S}/T_{2S}$, as a function of NMR frequency $f_0$ and heptane concentration $\phi_{C7}$ in the polymer-heptane mix. Our MD simulations of $T_{1}$, $T_{2}$ and diffusion are also compared to previously reported measurements of similar systems \cite{singer:EF2018}.

The rest of this manuscript is organized as follows: section \ref{sc:Methodology} presents the methodology for the polymer-heptane mixtures, MD simulations of diffusion and relaxation, auto-correlation and NMR relaxation, and effects of dissolved oxygen on measurements; followed by the results in section \ref{sc:Results} for diffusion of heptane in the mix, intra and inter-molecular relaxation, total relaxation, surface relaxation and relaxivity of heptane in the polymer matrix; followed by the conclusions in section \ref{sc:Conclusions}.

\section{Methodology}\label{sc:Methodology}

\subsection{Polymer-heptane mixtures}

\begin{figure}[ht!]
\centering
\includegraphics{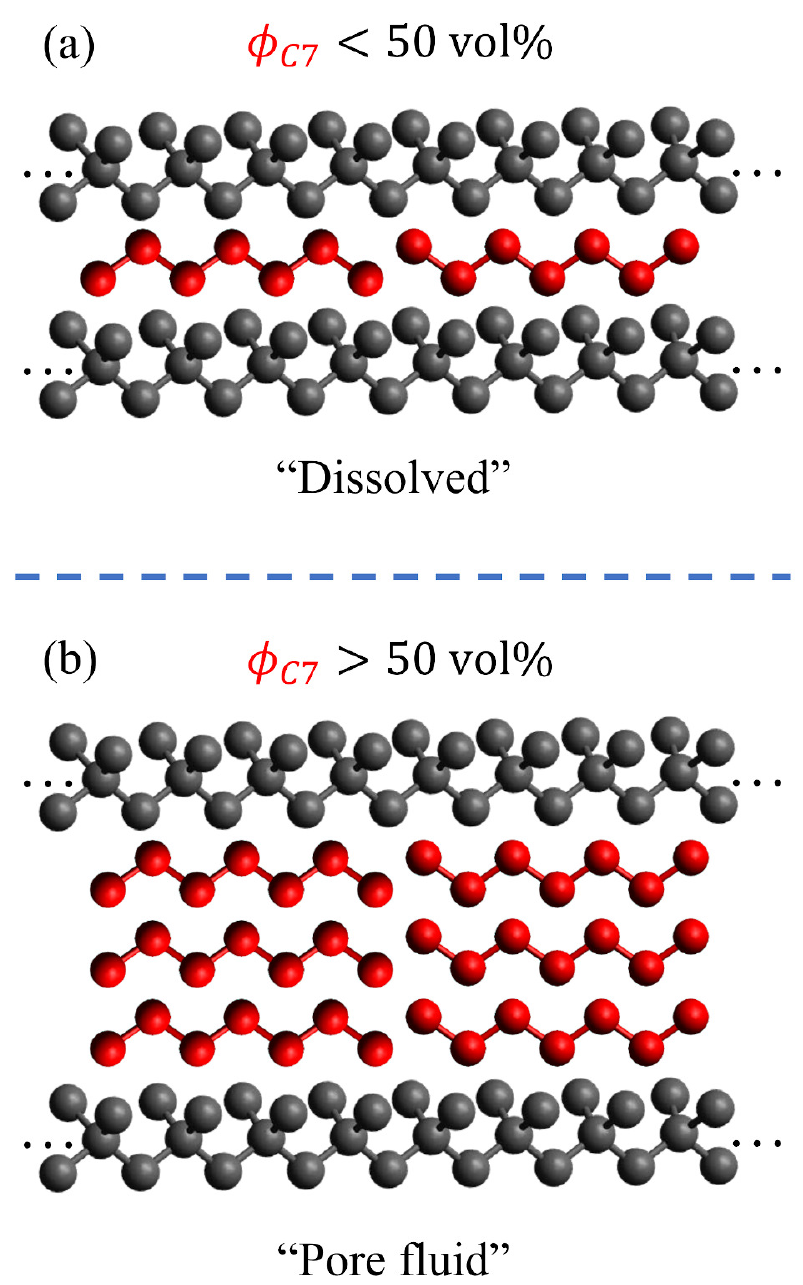}
\caption{Illustration of a cross-section of (locally) cylindrical transient ``pores" in a poly(isobutene) matrix (black) filled with $n$-heptane (red), where only carbon atoms are shown. x(a) ``Dissolved" regime corresponds to simulations with low heptane volume fractions $\phi_{C7} < 50$ vol\% where heptane molecules rarely contact other heptane molecules. (b) ``Pore fluid" regime corresponds to simulations with high heptane volume fractions $\phi_{C7} > 50$ vol\% where heptane molecules fill a more conventional pore.} \label{fg:Porefluid}
\end{figure}

An illustration of the polymer-heptane mix is shown in Fig. \ref{fg:Porefluid}, where $n$-heptane is used as the representative alkane, and a 16-mer oligomer of poly(isobutene) of molecular mass $M_w = $ 912 g/mol is used for the polymer matrix. Poly(isobutene) is based on a Brookfield viscosity standard used in previously reported NMR measurements of polymer-heptane mixes, where the empirical relation $\eta \simeq A\, M_w^{\alpha}$ was reported with $\alpha \simeq 2.4 $ and $A \simeq 1.07 \times 10^{-4}$ at ambient temperature \cite{singer:EF2018}. The empirical viscosity relation implies that the simulated poly(isobutene) 16-mer has a viscosity of $\eta \simeq $ 1000 cP at ambient. Note that $M_w$ and $\eta$ for the simulations are lower than the measured polymer where $M_w$ = 9436 g/mol and $\eta \simeq  $ 333$\,$400 cP at ambient temperature and pressure \cite{singer:EF2018}. 

Fig. \ref{fg:Porefluid}(a) shows the ``dissolved" regime corresponding to simulations with low heptane volume fractions $\phi_{C7} < 50$ vol\%, where heptane molecules rarely contact other heptane molecules. Fig. \ref{fg:Porefluid}(b) shows the ``pore fluid" regime corresponding to simulations with high heptane volume fractions $\phi_{C7} > 50$ vol\% where heptane molecules fill a more conventional pore.

\subsection{Simulation details}

Both $n$-heptane and the polymer are modeled using using the CHARMM General Force Field \cite{Vanommeslaeghe2009, cgenff}, which is known to accurately describe the thermophyscial properties as well as the NMR relaxation and diffusion properties\cite{singer:jmr2017} of hydrocarbons. To construct the simulation system, we first created the structure of $n$-heptane using Avogadro\cite{hanwell:JC2012,avogadro} and poly(isobutene) using the PRO-DRG server \cite{schuettelkopf:ACD2004,prodrg}. We then created $N$ copies of heptane and $M$ copies of poly(isobutene), and pack them separately at a low density (0.1 g/cm$^3$) using PACKMOL\cite{packmol:2009,packmol}. The initial numbers are chosen considering ideal mixing and using the experimental density of polymer (0.89 g/cm$^3$) \cite{singer:EF2018} and heptane (0.68 g/cm$^3$) \cite{nist} NIST at 298.15 K. The numbers are chosen in such a way that $100\%$ polymer corresponds to 40 molecules. The two boxes are then combined to form the initial simulation box. We use the NAMD\cite{NAMD,phillips:JCC2005} code to perform the simulations. The equations of motion are integrated using the Verlet algorithm with a time step of 1 fs.  
To remove possible steric clashes, we minimize the system energy using 1000 steps of conjugate gradient minimization. This starting system is necessarily at a much lower pressure due to the low density. We then compress the system to atmospheric pressure using Langevin dynamics, where the temperature of 298.15 K is controlled using a Langevin thermostat and the pressure of 1~atm.\ is controlled using a Langevin barostat. Compressing from a low density state also ensures we have a well mixed system. This is critical because relying on diffusive motion to ensure mixing is not recommended for systems with low diffusivity, such as the polymer-alkane melt.  

We find that after about 1~ns, all the systems studied here achieve a constant density and temperature. We equilibrate this system at constant temperature ($NVT$ ensemble) for 1 ns.  The temperature during this phase was controlled by reassigning velocities (obtained from a Maxwell-Boltzmann distribution) every 250 steps.  The subsequent production
run was carried out for 10 ns at constant NVE. Frames were archived every 100 steps for analysis. The Lennard-Jones interactions were smoothly switched to zero between from 13{\rm \AA} and 14~{\rm \AA}. We use the particle mesh Ewald procedure to describe electrostatic interactions, with a grid spacing of 0.5~{\rm \AA}.

\subsection{Diffusion coefficient}

The simulated diffusion coefficient is obtained using the Einstein relation
\begin{eqnarray}
D_{sim} = \frac{1}{6} \frac{\delta \langle \Delta r^2 \rangle_L}{\delta t} \, 
\end{eqnarray}
where $\langle \Delta r^2\rangle$ is the mean-square displacement of the center-of-mass of the molecule as a function of diffusion evolution time $t$. Following Yeh and Hummer \cite{Yeh2004} (see also Ref.~\citenum{singer:jmr2017}), we correct the simulated diffusion coefficient for finite size effects using the relation
\begin{eqnarray}
D_T = D_{sim}  + \frac{k_B T}{6 \pi \eta} \frac{\xi}{L}
\end{eqnarray}
where $D_T$ is the diffusion coefficient, $L$ is the simulation boxlength, $\eta$ is the shear viscosity, $T$ is the temperature of the system, $k_B $ is the Boltzmann constant, and $\xi = 2.837297$ is the Madelung constant. The viscosity
for the simulated polymer was estimated using correlations developed by Fox and Flory\cite{Fox1948,Fox1951} for mono-disperse poly(isobutene). The viscosity of the mixtures were obtained following correlations developed by McAllister\cite{McAllister1960}. The system-size correction term amounts to around $\sim7\%$ for high volume fractions of heptane, and reduces to  $\sim2\%$ at low fractions. 

\subsection{Auto-correlation and NMR relaxation}

\begin{figure}[!ht]
	\begin{center}
		\includegraphics[width=1\columnwidth]{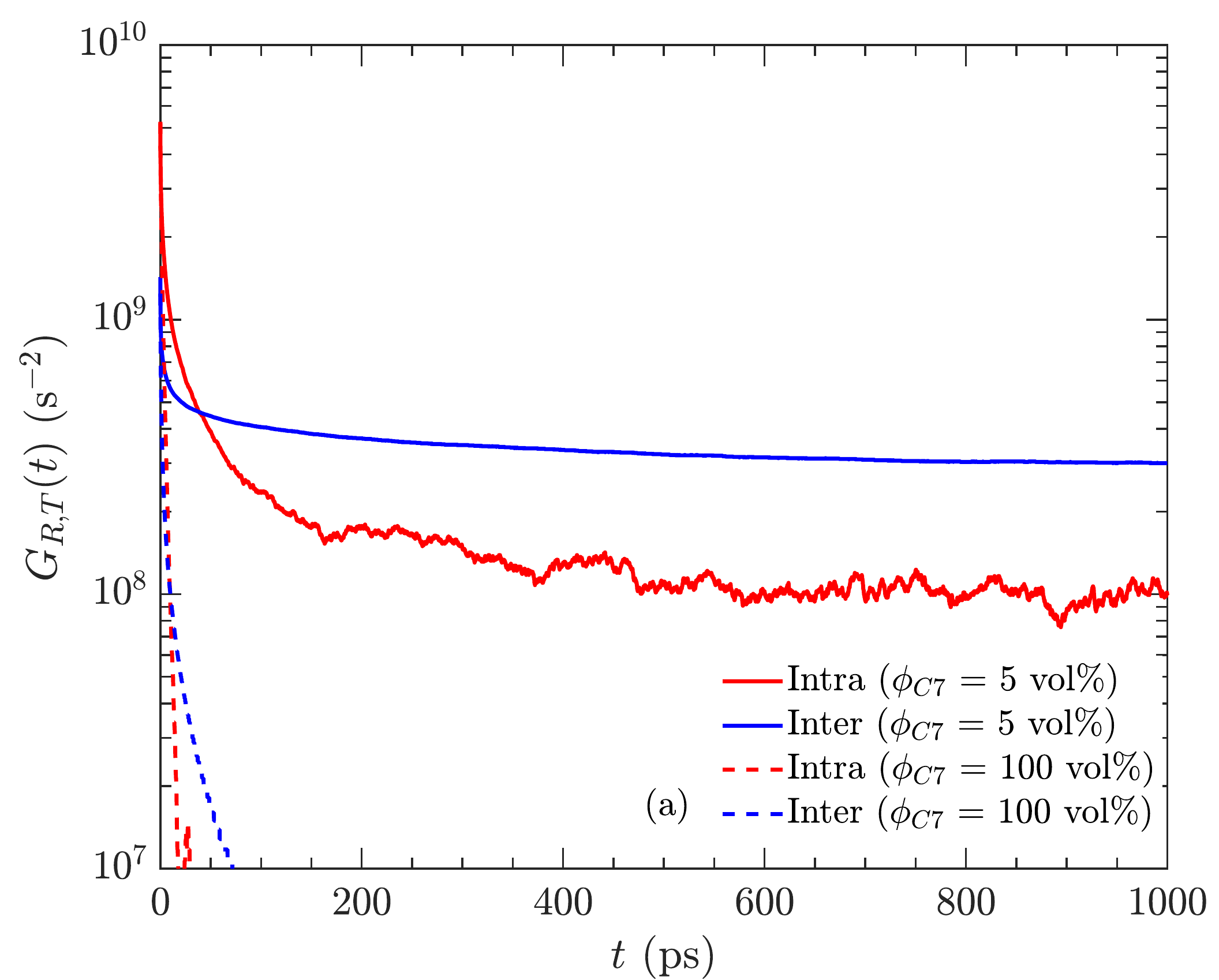} 
		\includegraphics[width=1\columnwidth]{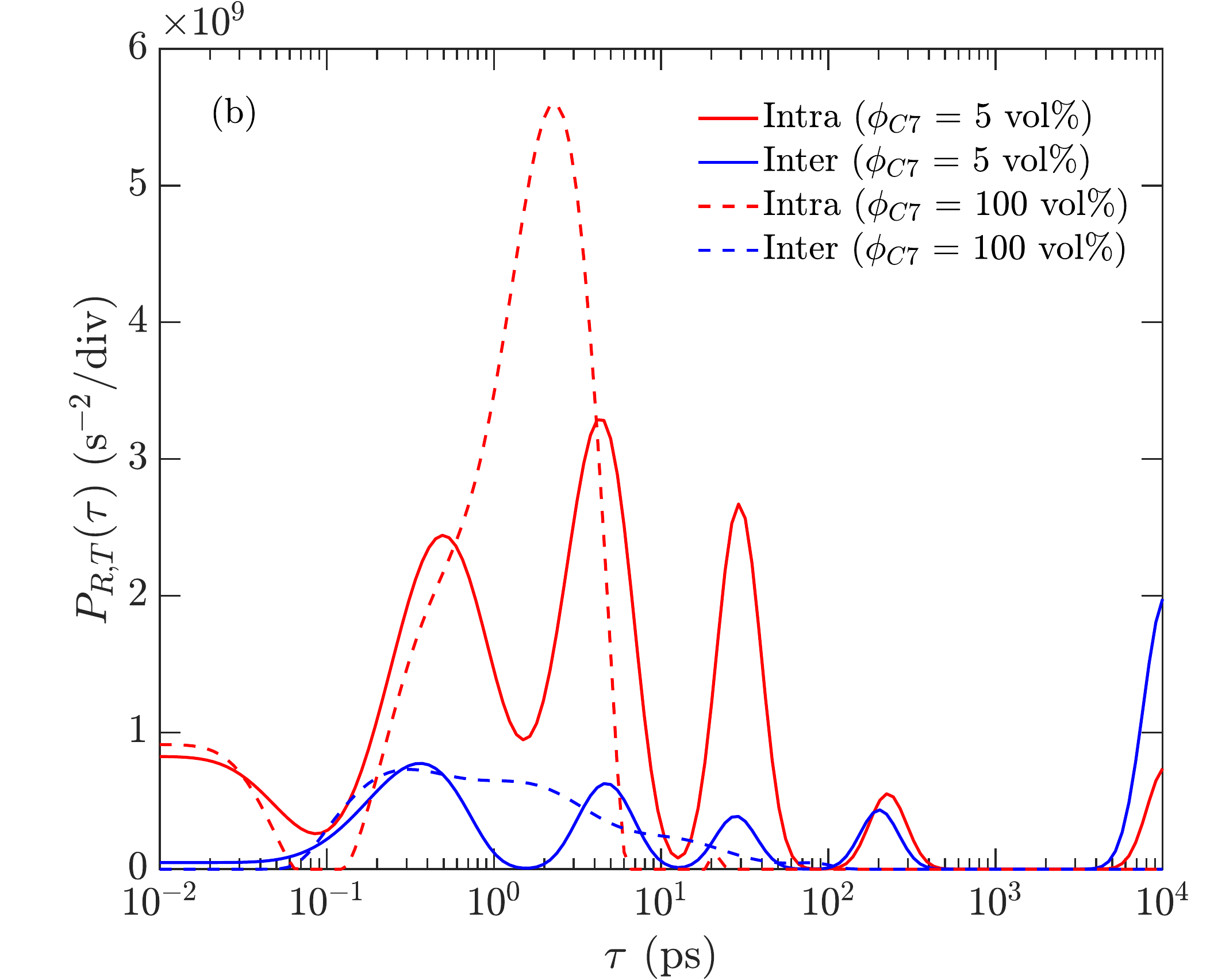} 
	\end{center}
	\caption{(a) MD simulations of the intra-molecular ($G_R(t)$) and inter-molecular ($G_T(t)$) auto-correlation functions for heptane in a polymer-heptane mix with heptane concentration $\phi_{C7}$ = 5 vol\%, compared with pure heptane ($\phi_{C7}$ = 100 vol\%). (b) Corresponding probability distributions $P_{R,T}(\tau)$ determined from inverse Laplace transforms of $G_{R,T}(t)$ (Eq. \ref{eq:ILT1}).} \label{fg:Gt}
\end{figure}

The autocorrelation function $G(t)$ for fluctuating magnetic $^1$H-$^1$H dipole-dipole interactions is central to the development of the NMR relaxation theory in liquids \cite{bloembergen:pr1948,torrey:pr1953,hwang:JCP1975,abragam:book,mcconnell:book,cowan:book,kimmich:book}. The details of the derivation of $G(t)$ can be found in Ref. \citenum{singer:jmr2017}, and only the essential elements are provided below. For an $isotropic$ system, $G(t)$ is given as: 
\begin{multline}
G_{R,T}(t) = \frac{3}{16} \! \left(\frac{\mu_0}{4\pi}\right)^2 \! \hbar^2 \gamma^4   \\
\times \frac{1}{N_{R,T}} \! \sum\limits_{i \neq j}^{N_{R,T}} \! \left< \frac{(3\cos^{2}\!\theta_{ij}\!(t+\tau)-1)}{r_{ij}^3\!\left(t+\tau\right)}  \frac{(3\cos^{2}\!\theta_{ij}\!(\tau)-1)}{r_{ij}^3\!(\tau)} \right>_{\!\! \tau}
\end{multline}
where $t$ is the lag time of the autocorrelation, $\mu_0$ is the vacuum permeability, $\hbar$ is the reduced Planck constant, $\gamma/2\pi = 42.58$ MHz/T is the nuclear gyro-magnetic ratio for $^1$H (spin $I=1/2$), $r_{ij}$ is the magnitude of the vector that connects the pair $(i,j)$ $^1$H-$^1$H dipoles, and $\theta_{ij}$ is the polar angle the vector forms with the external magnetic field. The subscript $R$ refers to autocorrelation of intra-molecular interactions from rotational diffusion, and subscript $T$ refers to autocorrelation of inter-molecular interactions from translational diffusion. From $G_{R,T}(t)$, one can determine the spectral density function $J_{R,T}(\omega)$ by Fourier transform as such:
\begin{equation}
J_{R,T}(\omega) = 2\int_{0}^{\infty}G_{R,T}(t)\cos\left(\omega t\right) dt,
\label{eq:FourierRTcos}
\end{equation}
for $G_{R,T}(t)$ in units of s$^{-2}$ \cite{mcconnell:book}. The expressions (which do not assume a molecular model) for $T_1$ and $T_2$  are then given by \cite{mcconnell:book,cowan:book}:
\begin{align}
\frac{1}{T_{1R,1T}} &= J_{R,T}(\omega_0) + 4 J_{R,T}(2\omega_0), \label{eq:T1RT} \\
\frac{1}{T_{2R,2T}} &= \frac{3}{2} J_{R,T}(0) + \frac{5}{2} J_{R,T}(\omega_0) + J_{R,T}(2\omega_0), \label{eq:T2RT}\\
\frac{1}{T_{1,2}} &= \frac{1}{T_{1R,2R}} + \frac{1}{T_{1T,2T}}, \label{eq:T12}
\end{align}
where $J_{R,T}(\omega_0)$ are the spectral densities at the resonance frequency $\omega_0 = 2\pi f_0$. Note that the intra-molecular and inter-molecular rates add to give the total relaxation rate (Eq. \ref{eq:T12}). 


The autocorrelation function $G_{R,T}(t)$ was constructed using fast Fourier transforms, for lag time ranging from $0$ ps to $1000$ ps in steps of $0.1$ ps. The number of distinct pairs of hydrogens analyzed are $\approx 2 \times 10^3$ per time frame for intra-molecular relaxation, and $\approx 4 \times 10^5$ per time frame for inter-molecular relaxation. 

The results of the intra-molecular $G_R(t)$ and inter-molecular $G_T(t)$ are shown in Fig. \ref{fg:Gt}(a) for heptane in the polymer-heptane mix at $\phi_{C7}$ = 5 vol\%, and for pure heptane (i.e. $\phi_{C7}$ = 100 vol\%). In order to quantify the departure of $G_{R,T}(t)$ from single-exponential decay, we fit $G_{R,T}(t)$ to a sum of multi-exponential decays and determine the underlying probability distribution $P_{R,T}(\tau)$ in correlation times $\tau$. More specifically, we perform an inversion of the following Laplace transform \cite{venkataramanan:ieee2002,song:jmr2002}: 
\begin{align}
G_{R,T}(t) &= \int_{0}^{\infty}\! P_{R,T}(\tau) \exp\left(-t/\tau\right) d\tau, \label{eq:ILT1}\\
\tau_{R,T} &= \frac{1}{G_{R,T}(0)}\int_{0}^{\infty} \! P_{R,T}(\tau)\, \tau \, d\tau,\label{eq:ILT2}\\
\Delta\omega^2_{R,T} &= 3 \,G_{R,T}(0), \textcolor{white}{\int_{0}^{\infty}}\label{eq:ILT3}\\
J_{R,T}(\omega) &= \int_{0}^{\infty}\! \frac{2\tau}{1+(\omega\tau)^2}   P_{R,T}(\tau) d\tau \label{eq:ILT4},
\end{align}
where $P_{R,T}(\tau)$ are the probability distribution functions derived from the inversion, and are plotted in Fig. \ref{fg:Gt}(b). Details of the inversion procedure can be found in \cite{singer:jcp2018b} and in the supporting information in \cite{singer:jcp2018}. The $P_{R,T}(\tau)$ in Fig. \ref{fg:Gt}(b) indicate a set of $\sim$5 polymer modes, located at similar $\tau$ values for both intra-molecular $P_{R}(\tau)$ and inter-molecular $P_{T}(\tau)$ interactions. The intra-molecular $P_{R}(\tau)$ has an additional mode at short $\tau\simeq10^{-2}$ ps for both the polymer and heptane, while it is absent for $P_{T}(\tau)$ in both cases. Similar observations at $\tau\simeq10^{-2}$ ps were reported in the supporting information in \cite{singer:jcp2018} for liquid-state alkanes.

\begin{figure}[!ht]
\centering
\includegraphics[width=1\columnwidth,trim=0 1.78cm 0 0, clip]{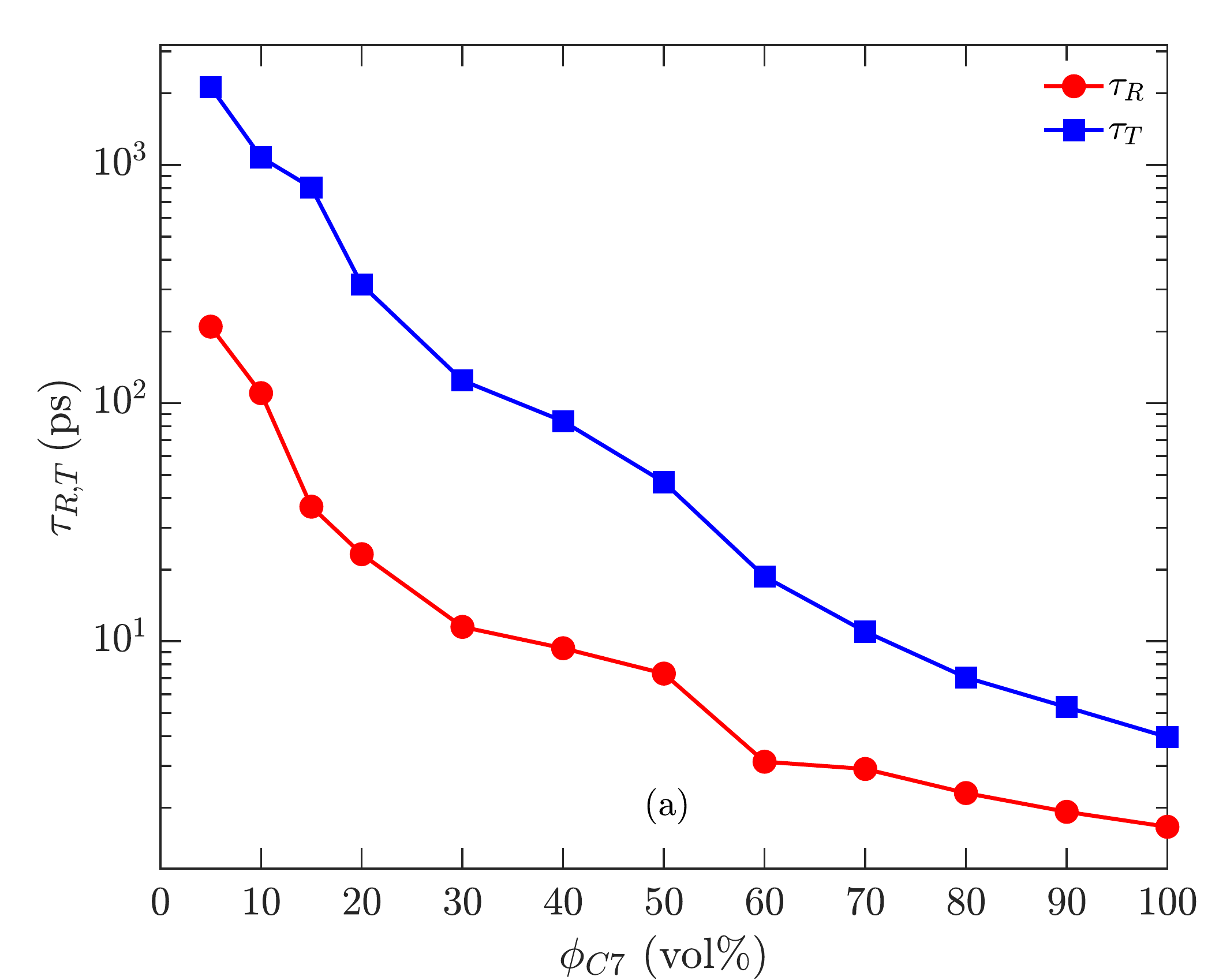} \includegraphics[width=1\columnwidth,trim=0 0 0 0.3cm, clip]{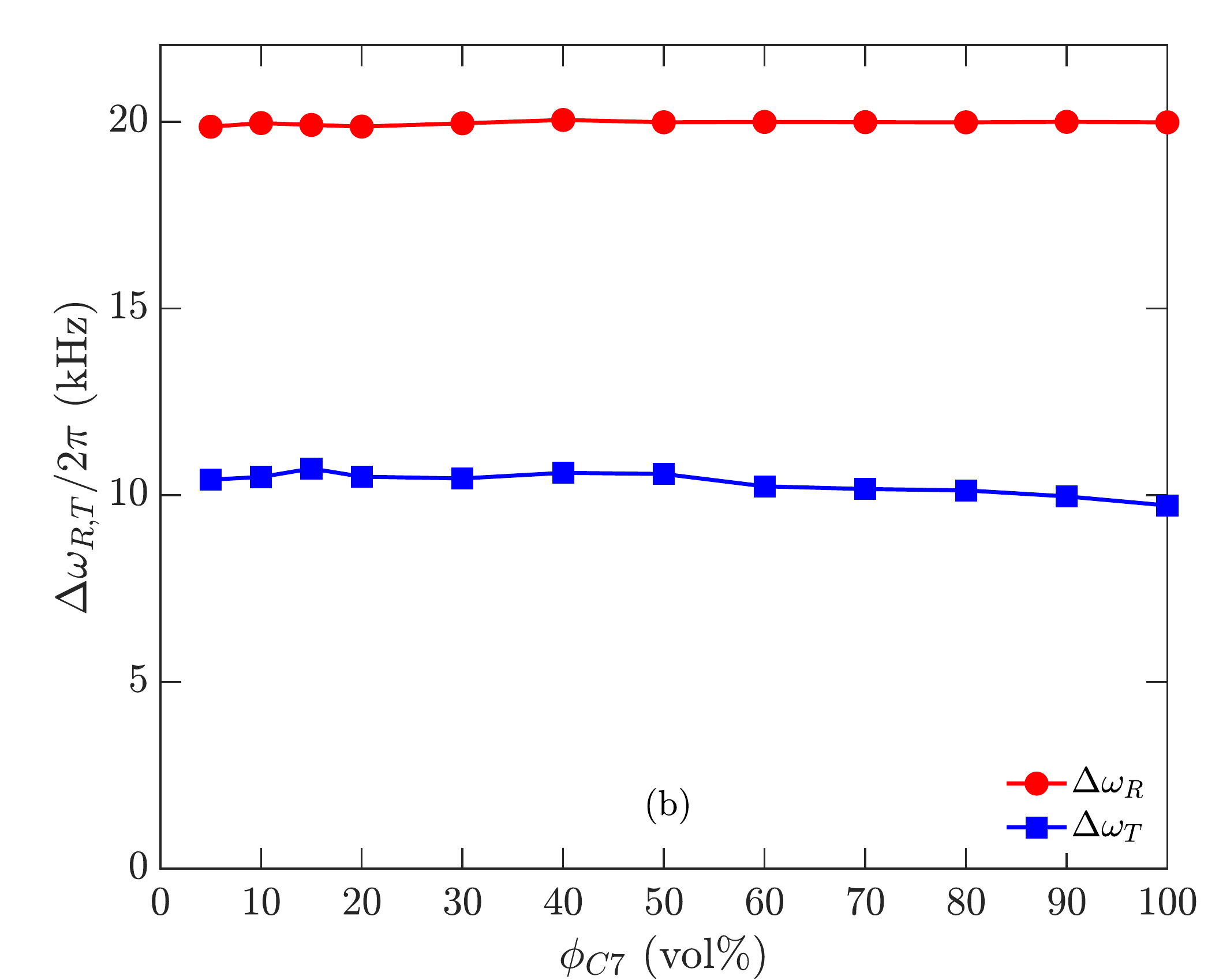}
\caption{(a) Correlation times for the rotational ($\tau_R(t)$) and translational ($\tau_T(t)$) motions as a function of heptane concentration $\phi_{C7}$. (b) Square-root of second moment (i.e. strength) of intra-molecular ($\Delta \omega_R$) and inter-molecular ($\Delta\omega_T$) interactions as a function of $\phi_{C7}$.}\label{fg:Tau}
\end{figure}

The decomposition of $G_{R,T}(t)$ into a sum of exponential decays in Eq. \ref{eq:ILT1} is common practice in phenomenological models of complex molecules \cite{beckmann:prep1988,bakhmutov:book}, where the more complex the molecular dynamics, the more exponential terms are required \cite{woessner:jcp1962,woessner:jcp1965}. Also defined are the correlation times $\tau_{R,T}$ (Eq. \ref{eq:ILT2}), and the square-root of the second moments $\Delta\omega_{R,T}$ (Eq. \ref{eq:ILT3}) i.e. strength of the interaction, which are plotted in Fig. \ref{fg:Tau}. 
While $\Delta\omega_{R,T}$ are independent of $\phi_{C7}$, $\tau_{R,T}$ increases by $\sim$3 orders of magnitude in going from $\phi_{C7} $ = 100 vol\% $\rightarrow$ 5 vol\%. The increase in $\tau_{R,T}$ clearly show that decreasing $\phi_{C7} $ dramatically slows the molecular dynamics of heptane due to increasing confinement in the polymer matrix.

$T_{1,2}$ as a function of $f_0$, i.e. $T_{1,2}$ dispersion, can also be determined from the $P_{R,T}(\tau)$ distributions. This is derived by using the Fourier transform (Eq. \ref{eq:FourierRTcos}) of $G_{R,T}(t)$ (Eq. \ref{eq:ILT1}), resulting in Eq. \ref{eq:ILT4}. Once $J_{R,T}(\omega)$ is known, Eqs. \ref{eq:T1RT}, \ref{eq:T2RT} and \ref{eq:T12} are used to determine $T_{1,2}(\omega_0)$ from $J_{R,T}(\omega)$ at $\omega = \omega_0$. 

\section{Results}\label{sc:Results}

\subsection{Diffusion of heptane in the mix}

\begin{figure}[!t]
	\centering
	\includegraphics[width=1\columnwidth]{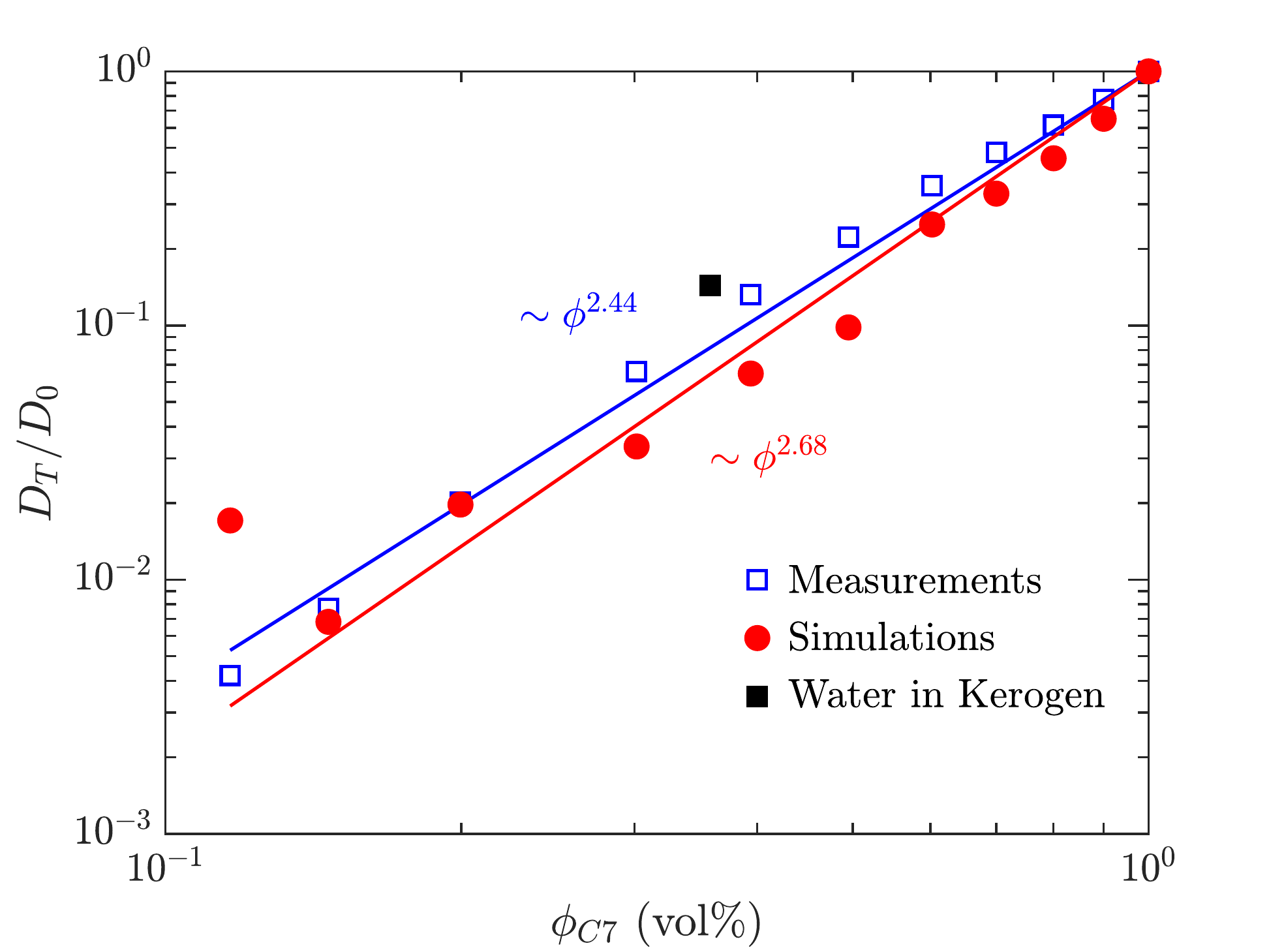}
	\caption{Ratio of translational diffusion coefficient $D_T$ to bulk diffusion coefficient $D_0$ for heptane in the polymer matrix for both measurements and simulations, as a function of heptane concentration $\phi_{C7}$. Lines are best fits using a Archie model Eq. \ref{eq:Archie} for tortuosity. Also shown are measurements of restricted diffusion of water in immature Kerogen, see  Appendix \ref{SI:DT2expt} for more detail.}
	\label{fg:Diff}
\end{figure}

Fig.~\ref{fg:Diff} compares the simulated diffusion coefficient against NMR diffusion measurements in the polymer-heptane mixtures. We emphasize that the polymer matrix used experimentally is of considerably higher viscosity ($\eta \simeq$ 333$\,$400 cP at ambient) than the one used in simulations ($\eta \simeq$ 1000 cP at ambient). Nevertheless, both simulations and measurements show that the diffusion coefficient $D_T$ relative to the value in the bulk $D_0$ (= 3.43$\times$10$^{-9}$ m$^2$/s at ambient) is consistent with \cite{dunn:book2002}:
\begin{align}
\frac{D_T}{D_0} &= \frac{1}{\mathcal T} = \phi_{C7}^{m-1}, \label{eq:Archie}
\end{align}
where ${\mathcal T}$ is the tortuosity, and ${\mathcal T} = \phi_{C7}^{1-m}$ is Archie's equation with cementation exponent $m$. The simulations indicate that $m\simeq 3.68$, which agrees well with NMR measurements where $m\simeq 3.44$. By comparison, $m$ = 2 is predicted in a capillary bundle model \cite{dunn:book2002}. It should be noted however that in the present case the polymer is itself diffusing, and therefore the pore walls are not rigid. 

We repeated our calculations for the low volume fractions ($\leq$ 30\%) for reproducibility, as we expected biases from our initial configurations. For the lowest volume fractions ($\leq$ 10\%), we find the values can vary by a factor of about $\sim$5. This serves as a cautionary note in the challenges that remain in simulating systems with very low diffusivity. However, the overall agreement with measurements is encouraging, which gives us confidence in our molecular models and simulation forcefield. 

Furthermore, Fig.~\ref{fg:Diff} also shows that the results for heptane in the polymer-heptane mix are consistent with previously reported NMR measurements of water in immature kerogen isolates. This suggests that the high-viscosity polymer is a good model for translational diffusion of fluids in an immature kerogen matrix, which is reasonable given that immature kerogen has fewer cross-links than mature kerogen \cite{liu:EF2019}.

\subsection{Intra- vs. inter relaxation of heptane}

\begin{figure}
\centering
\includegraphics[width=1\columnwidth,trim=0 1.78cm 0 0, clip]{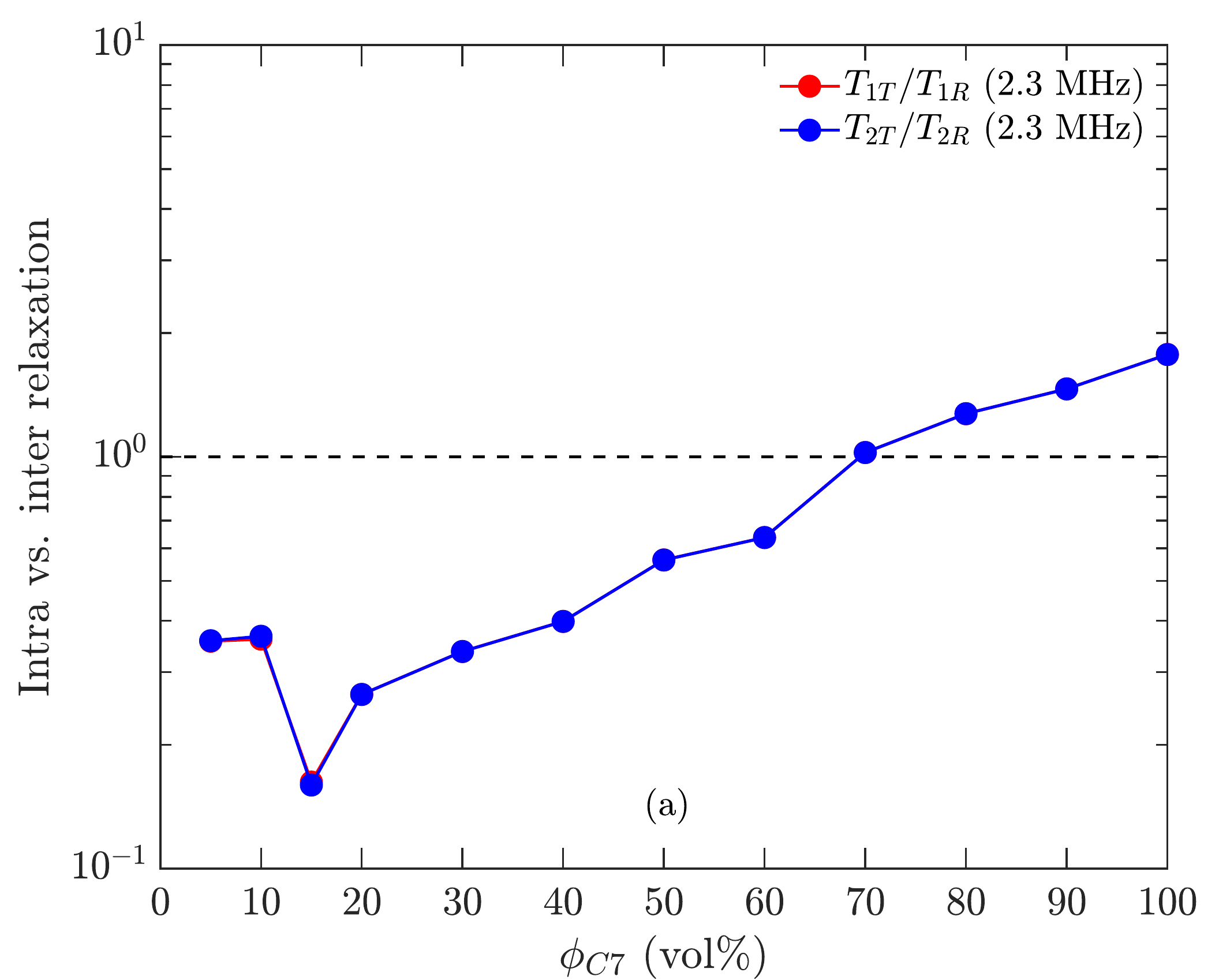} \includegraphics[width=1\columnwidth,trim=0 1.78cm 0 0.3cm, clip]{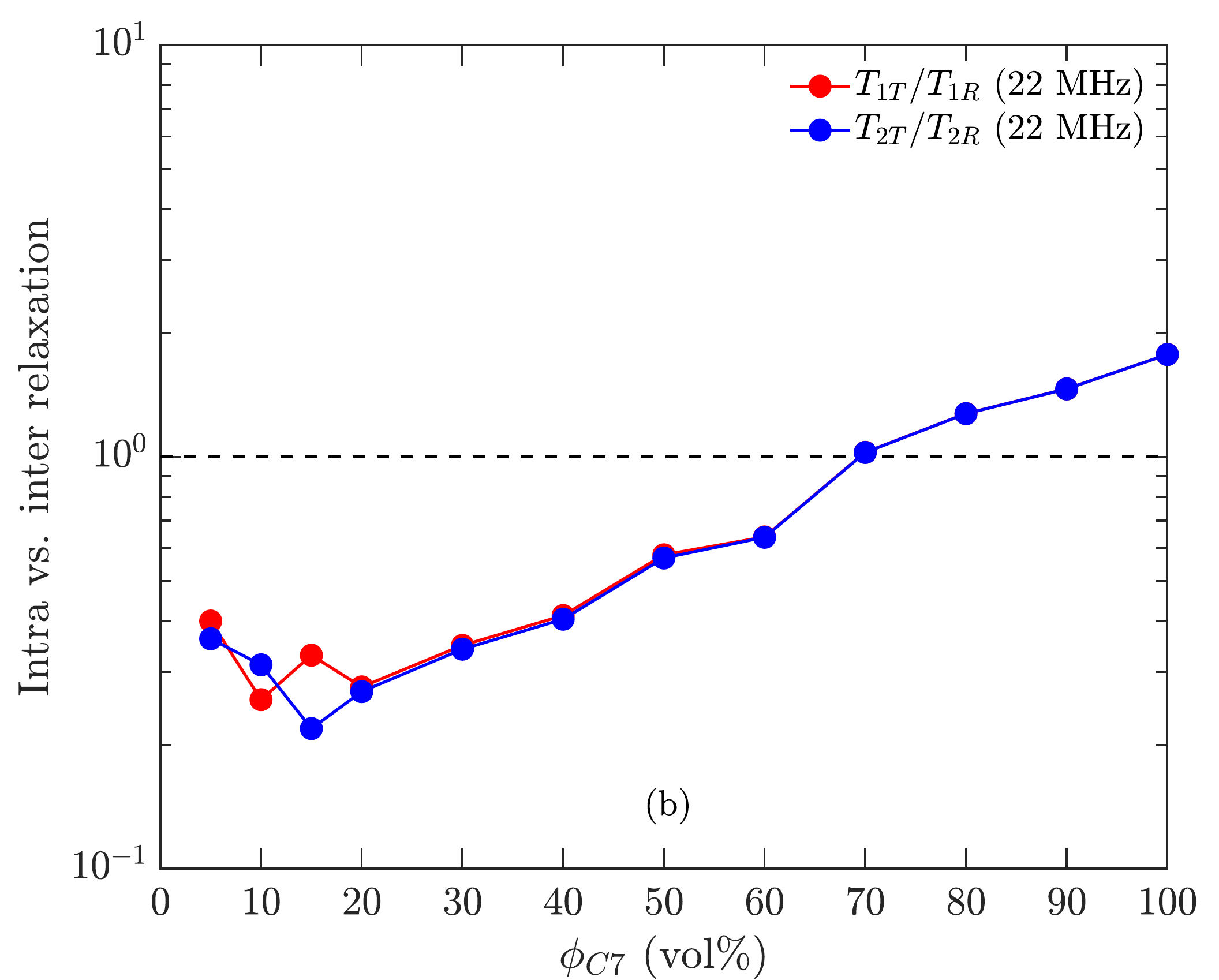} \includegraphics[width=1\columnwidth,trim=0 0 0 0.3cm, clip]{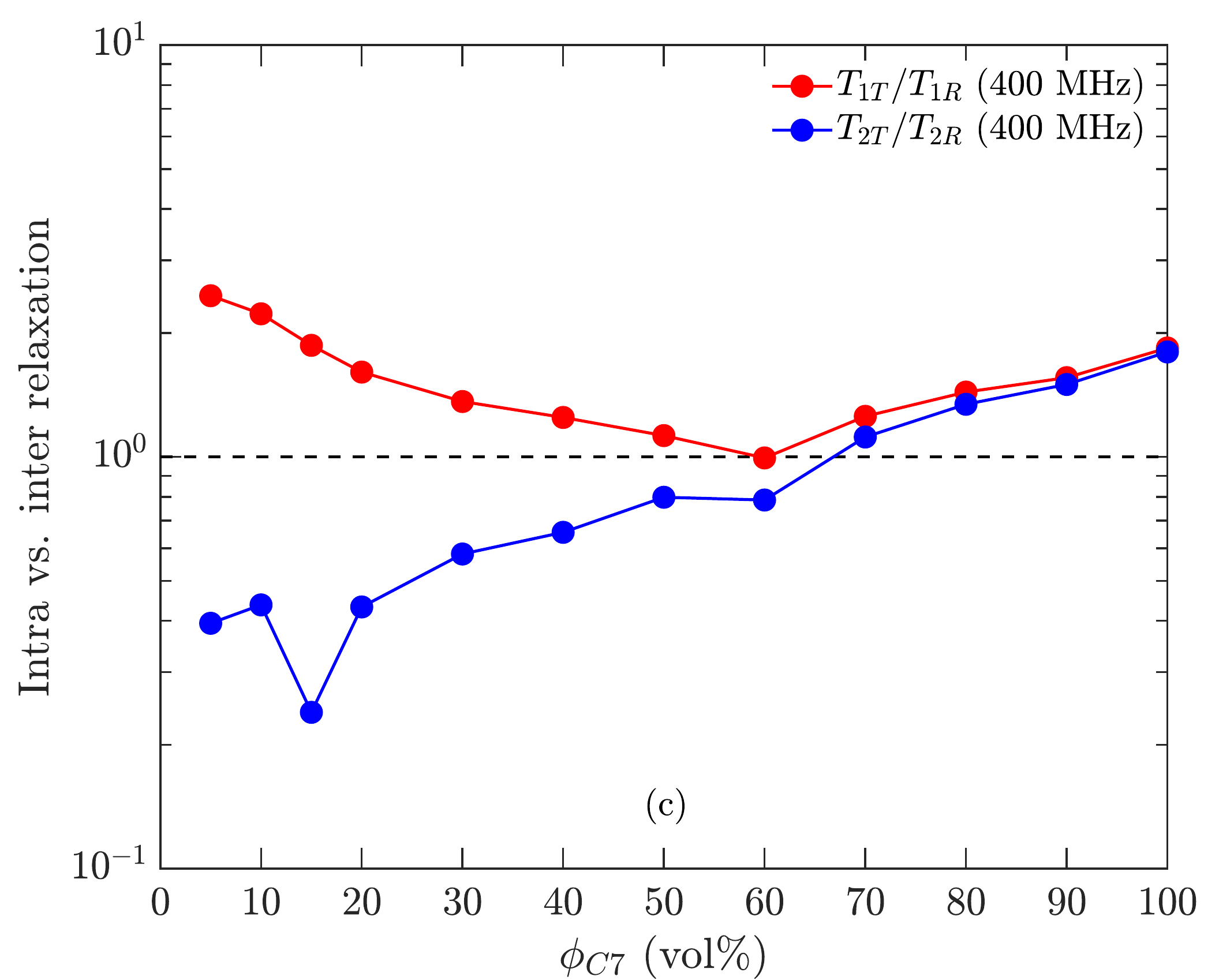} 
\caption{Ratios of inter-molecular ($T_{1T,2T}$) to intra-molecular ($T_{1R,2R}$) relaxation times at $f_0$ = (a) 2.3 MHz, (b) 22 MHz, and (c) 400 MHz, as a function of heptane concentration $\phi_{C7}$. Ratios greater than one (dashed lines) indicate that intra-molecular relaxation dominates over inter-molecular relaxation, while ratios less than one indicate the opposite.}
\label{fg:Ratios}
\end{figure}

MD simulations can naturally separate intra-molecular $T_{1R,2R}$ from inter-molecular $T_{1T,2T}$ relaxation. In Fig. \ref{fg:Ratios} we show the ratios of $T_{1T}/T_{1R}$ and $T_{2T}/T_{2R}$, where ratios greater than one indicate that intra-molecular relaxation dominates over inter-molecular relaxation, while ratios less than one indicate the opposite. It is found that intra-molecular relaxation dominates over inter-molecular (i.e. $T_{1T,2T}/T_{1R,2R} > 1$) for $\phi_{C7} \gtrsim $ 70 vol\%. Below $\phi_{C7} \lesssim $ 70 vol\%, inter-molecular relaxation dominates (i.e. $T_{1T,2T}/T_{1R,2R} < 1$), except at high frequencies $f_0 \gtrsim $ 400 MHz where the reverse is found for $T_{1T}/T_{1R} > 1$. 

The dependence of $T_{2T}/T_{2R}$ on $\phi_{C7} $ can be loosely described from the results in Fig. \ref{fg:Tau} alone. According to Eq. \ref{eq:T2RT}, the $1/T_{2R,2T} \propto \Delta\omega_{R,T}^2 \tau_{R,T}$, if one ignores the small amount of dispersion in $T_{2R,2T}$. At $\phi_{C7} = $ 100 vol\%, $\Delta\omega_{R}^2 $ is a factor $\sim$4 larger than $\Delta\omega_{T}^2 $, while $\tau_R$ is a factor $\sim$1/2 shorter than $\tau_T$, which leads to $T_{2T}/T_{2R} \simeq 2$ (i.e. intra-molecular relaxation dominates). With decreasing $\phi_{C7} $, $\Delta\omega_{R,T}^2 $ stay constant (Fig. \ref{fg:Tau}(b)), however $\tau_T$ becomes larger than $\tau_R$ (Fig. \ref{fg:Tau}(a)), which leads to $T_{2T}/T_{2R} < 1$ (i.e. inter-molecular relaxation dominates) below $\phi_{C7} \lesssim $ 70 vol\%. Note that inter-molecular relaxation consists of contributions from both heptane-heptane interactions and polymer-heptane interactions. However below $\phi_{C7} \lesssim $ 50 vol\%, i.e. in the dissolved region (Fig. \ref{fg:Porefluid}(a)), heptane molecules do not contact other heptane molecules, therefore the inter-molecular relaxation is dominated by polymer-heptane interactions.

The dependence of $T_{1T}/T_{1R}$ on $\phi_{C7} $ can be understood using $P_{R,T}(\tau)$ in Fig. \ref{fg:Gt}(b). According to Eq. \ref{eq:ILT4}, contributions from $\omega\tau \gg 1$ are negligible compared to those with $\omega\tau \ll 1$. Contributions with $\omega\tau \gg 1$ are often said to be ``dispersed out" \cite{caravan:cr1999}, meaning they do not contribute to relaxation compared to $\omega\tau \ll 1$. For $\phi_{C7} = $ 5 vol\% at $f_0 = $ 400 MHz this implies that contributions from $P_{R,T}(\tau)$ with $\tau \gg 400 $ ps are negligible. In other words, for $\phi_{C7} = $ 5 vol\% in Fig. \ref{fg:Gt}(b), the peak at $\tau \simeq 10^4$ ps (where the inter-molecular contribution is larger than intra-molecular contribution) no longer contributes to relaxation $f_0 = $ 400 MHz, i.e. it is dispersed out. Meanwhile $P_{R}(\tau)$ and $P_{T}(\tau)$ have peaks at similar $\tau$ values for $\tau \ll 400 $ ps, however $P_{R}(\tau)$ is a factor $\sim$4 larger in amplitude than $P_{T}(\tau)$ in that region, therefore $T_{1T}/T_{1R} > 1$ (i.e. intra-molecular relaxation dominates). At lower frequencies $f_0 \lesssim $ 400 MHz, the peak at $\tau \simeq 10^4$ ps in Fig. \ref{fg:Gt}(b) is not dispersed out, therefore similar arguments for $T_{1T}/T_{1R}$ hold as for $T_{2T}/T_{2R}$ above, where $T_{1T}/T_{1R} < 1$ below $\phi_{C7} \lesssim $ 70 vol\% (i.e. inter-molecular relaxation dominates). 


\subsection{Total relaxation of heptane}

\begin{figure}
\centering
\includegraphics[width=1\columnwidth,trim=0 1.78cm 0 0, clip]{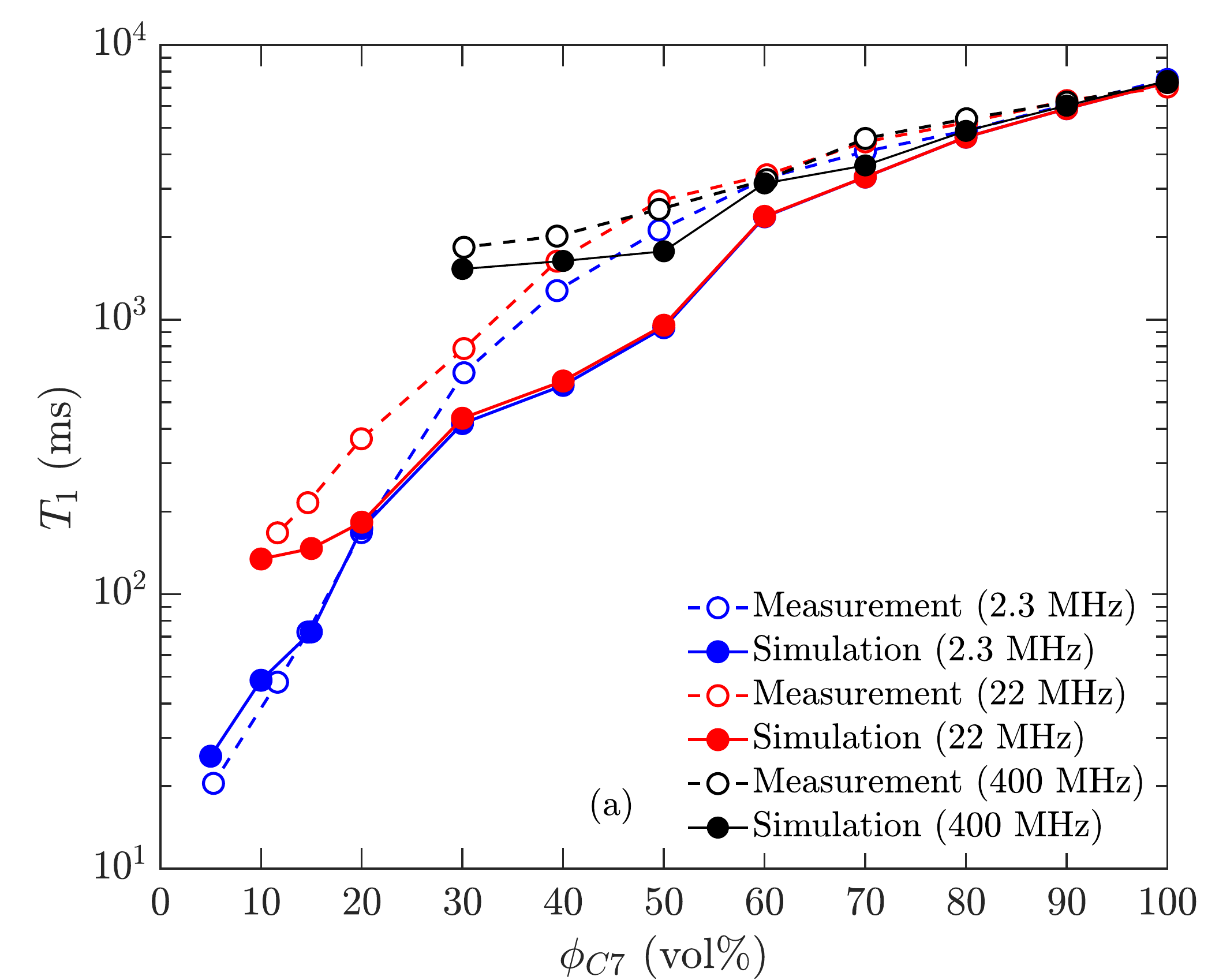} \includegraphics[width=1\columnwidth,trim=0 0 0 0.3cm, clip]{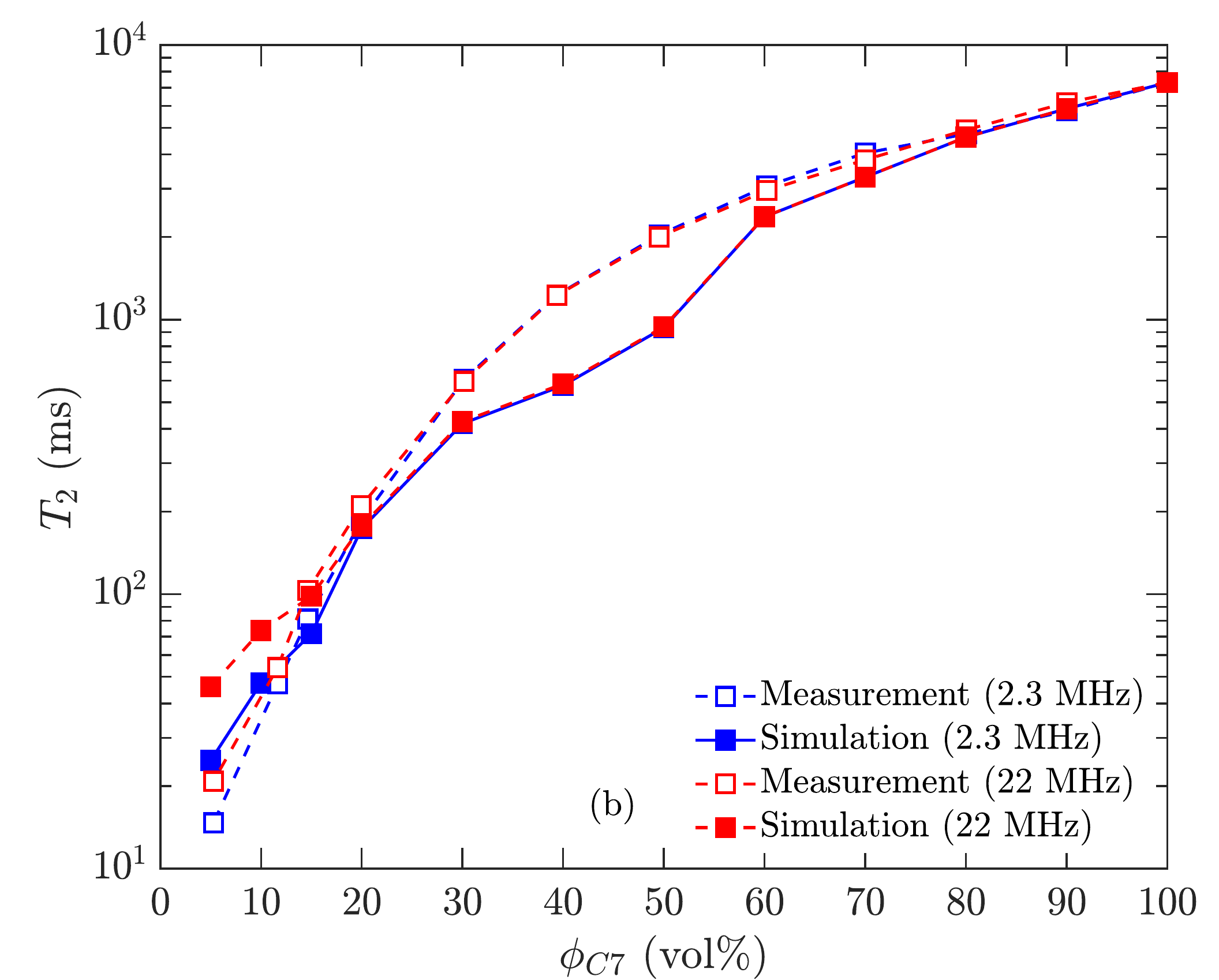}
\caption{(a) $T_{1}$ and (b) $T_2$ relaxation times for heptane in the polymer matrix according to simulations (closed symbols) and measurements (open symbols) at $f_0$ = 2.3 MHz, 22 MHz, 400 MHz, as a function of heptane concentration $\phi_{C7}$.}
\label{fg:T1T2both}
\end{figure}

The expression for the total relaxation is given in Eq. \ref{eq:T12}, and is plotted in Fig. \ref{fg:T1T2both} as a function of $\phi_{C7}$ at $f_0$ = 2.3 MHz, 22 MHz, 400 MHz. $T_2$ decreases monotonically with decreasing $\phi_{C7}$, which as shown below is a result of larger surface relaxation due to increased confinement (i.e. a larger surface to pore-volume ratio), and a larger surface relaxivity $\rho_2$ in the dissolved region. Likewise, $T_1$ decreases monotonically with decreasing $\phi_{C7}$ at $f_0$ = 2.3 MHz and 22 MHz. However at $f_0$ = 400 MHz, $T_1$ tends to level off with decreasing $\phi_{C7}$, which as shown below is due to a constant surface-relaxivity $\rho_1$ in the dissolved region.

\begin{figure}
\centering
\includegraphics[width=1\columnwidth]{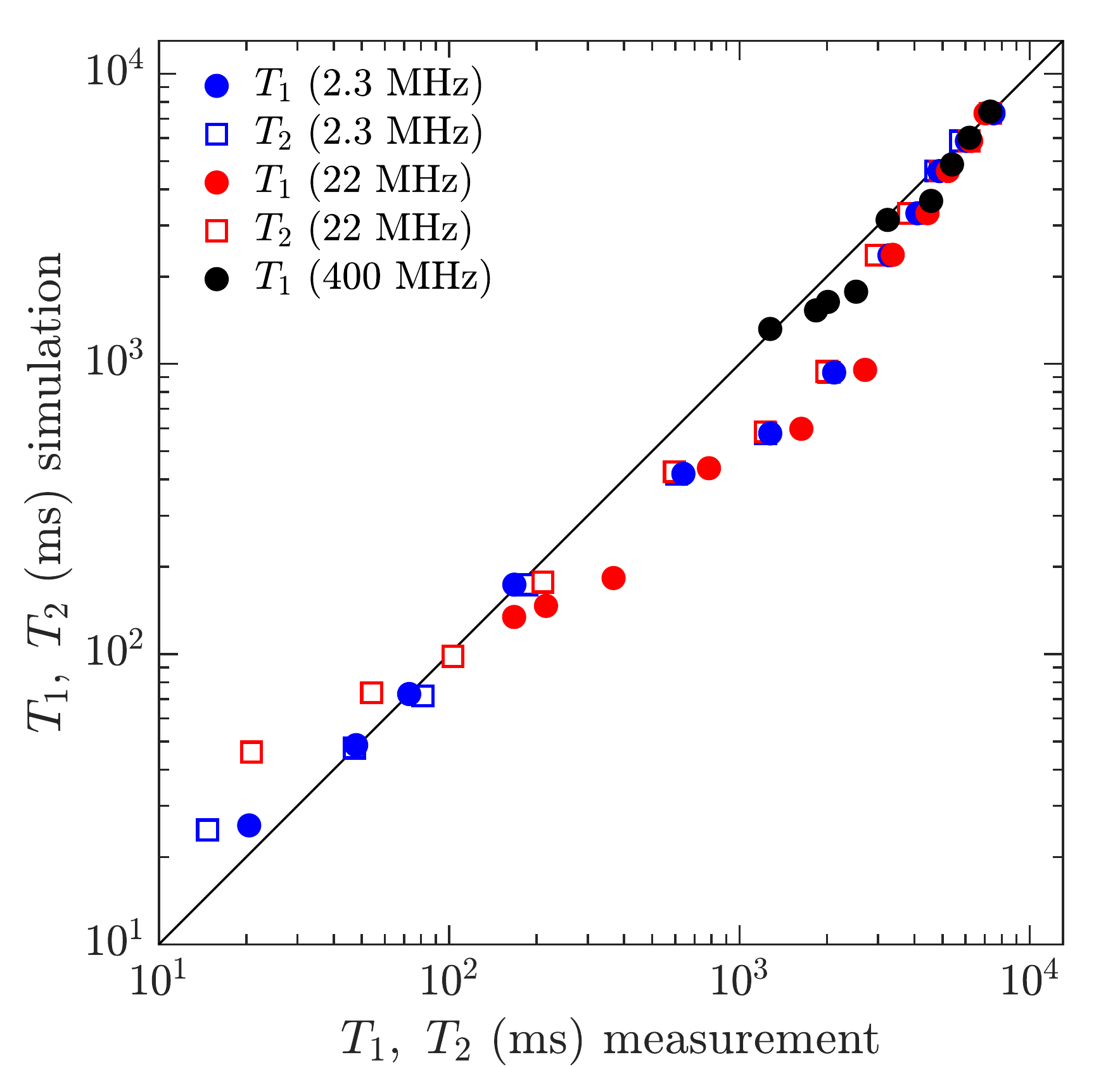} 
\caption{Correlation cross-plot of measurements vs. simulations of $T_1$ and $T_2$ at $f_0$ = 2.3 MHz, 22 MHz, 400 MHz, for various heptane volume fractions $\phi_{C7}$.}
\label{fg:Crossplot}
\end{figure}

Also shown in Fig. \ref{fg:T1T2both} are the measurements compared with simulations, which shows that the overall trends agree. A cross-plot of measurements versus simulations is also shown in Fig. \ref{fg:Crossplot} for better comparison. We find good agreement between the measurements and simulations except in two regions: (1) the region $\phi_{C7} \simeq $ 50 vol\% where the measurements are overestimated compared to simulations, and (2) the region $\phi_{C7} \lesssim $ 10 vol\% where the simulations are overestimated compared to measurements. 

The deviation in the region $\phi_{C7} \lesssim $ 10 vol\% is most likely due to the fact that the maximum auto-correlation time for $G_{R,T}(t)$ in Fig. \ref{fg:Gt}(a) is $t_{max} = $ 1000 ps, which limits the accuracy in the inverse Laplace transforms $P_{R,T}(\tau)$ for $\phi_{C7} \lesssim $ 10 vol\%. The maximum value of $\tau_{max}$ in $P_{R,T}(\tau)$ is chosen to be a factor $\sim$10 larger than the longest acquisition time in the data \cite{venkataramanan:ieee2002}, i.e. $\tau_{max} = 10\, t_{max} = 10^4$ ps in the present case. This leads to inaccuracies in $P_{R,T}(\tau)$ if there are contributions with $\tau > \tau_{max}$, which is likely to be the case for $\phi_{C7} \lesssim $ 10 vol\%. The solution is to increase $t_{max} $, however this is computationally expensive.

The deviation in the region $\phi_{C7} \simeq $ 50 vol\% is believed to be due to uncertainties in the oxygen concentration in the measurements. This is explored in Appendix \ref{SI:O2conc}, and the proposition only qualitatively improves the agreement with experiments.

\subsection{Surface relaxation of heptane}

The decrease in heptane relaxation times $ T_{1,2}$ from their bulk relaxation $ T_{1B,2B}$ is due to interactions of heptane with the polymer surfaces. Commonly, this property is analyzed as relaxation induced by the surface itself and termed ``surface relaxation". As indicated in Fig.~\ref{fg:Porefluid}, if we assume the polymer matrix to form ``pores" for the heptane molecules, we can interpret the polymer-heptane interactions as ``surface" interactions, where the polymer is the confining surface.

\begin{figure}[h!]
\centering
\includegraphics[width=1\columnwidth,trim=0 1.78cm 0 0, clip]{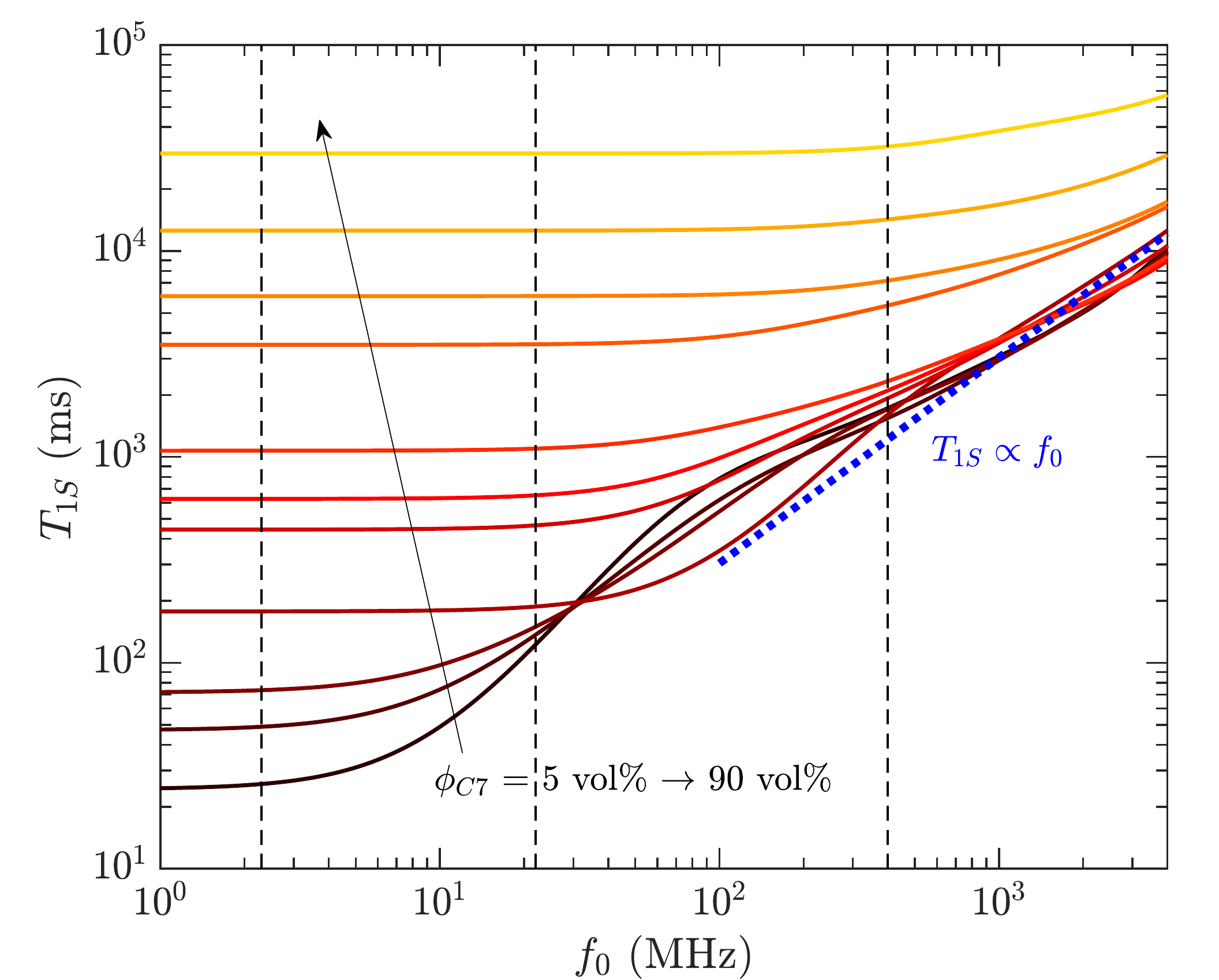} \includegraphics[width=1\columnwidth,trim=0 1.78cm 0 0.3cm, clip]{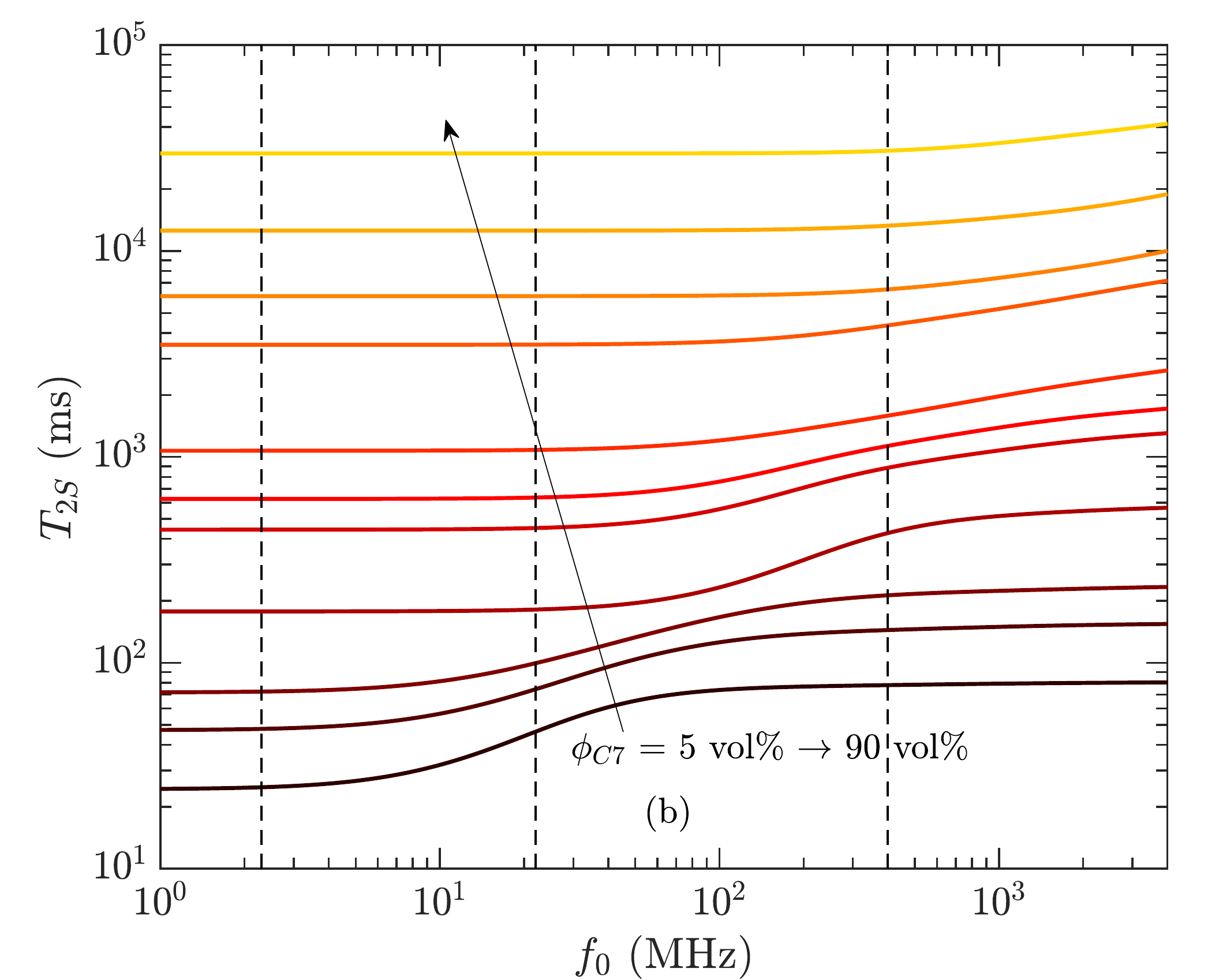}
 \includegraphics[width=1\columnwidth,trim=0 0 0 0.3cm, clip]{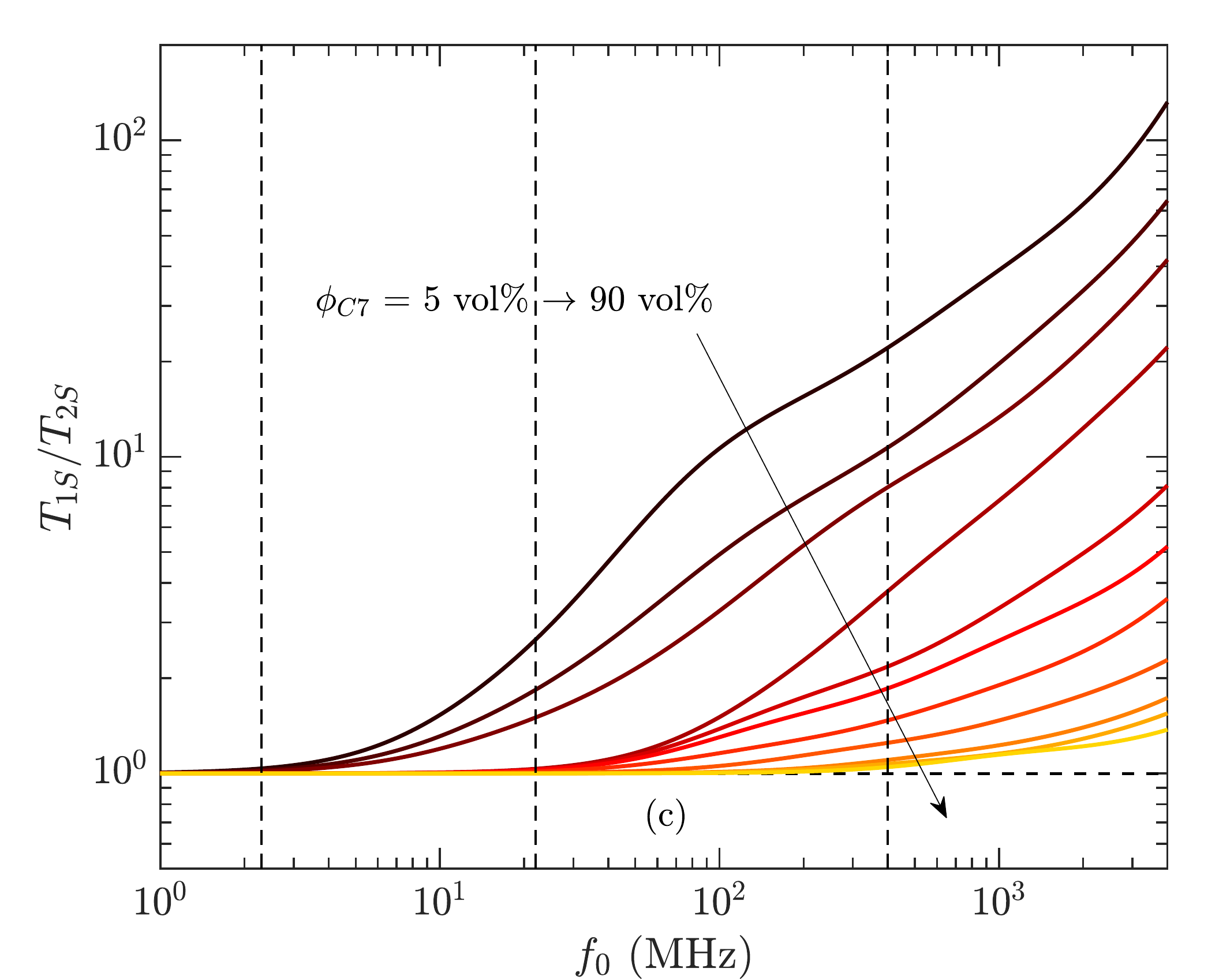} 
\caption{Surface relaxations (a) $T_{1S}$, (b) $T_{2S}$, and (c) $T_{1S}/T_{2S}$ ratio as a function of frequency $f_0$, for heptane volume fraction $\phi_{C7}$ of 5\%, 10\%, 15\%, 20\% to 90\% in increments of 10\%. Dashed blue line in (a) shows dispersion relation $T_{1S}\propto f_0$ (specifically $T_{1S} \times 2.3/f_0 = 7$ ms).}
\label{fg:T1ST2S}
\end{figure}

The surface relaxation is obtained using the following relation: 
\begin{eqnarray}
\frac{1}{T_{1,2}} &=& \frac{1}{T_{1S,2S}} + \frac{1}{T_{1B,2B}},
\end{eqnarray}
where $T_{1B,2B} \simeq$ 7320 ms  is the bulk relaxation time for heptane at ambient temperatures ($\eta$ = 0.39 cP) \cite{lo:SPE2002,shikhov:amr2016}. Fig. \ref{fg:T1ST2S} shows the resulting dispersion for $T_{1S}$, $T_{2S}$, and the ratio $T_{1S}/T_{2S}$, for the various $\phi_{C7}$ mixtures. 

While $T_{1B,2B}$ for bulk heptane has minimal frequency dependence within the range shown (simulation not shown), $T_{1S}$ for heptane in the polymer matrix clearly has a large amount of dispersion. More specifically, the simulations show that $T_{1S}$ is dispersive above $f_0 \gtrsim $ 10 MHz, and furthermore that $T_{1S}$ tends towards the functional form $T_{1S}\propto f_0$ (specifically $T_{1S} \times 2.3/f_0 \simeq 7$ ms) for $\phi_{C7} \lesssim $ 50 vol\% and high frequencies $f_0 \gtrsim $ 500 MHz. This functional form for $T_{1S}$ dispersion is consistent with the previously reported measurements of polymers and bitumen where $T_{1LM}\propto f_0$ (specifically $T_{1LM} \times 2.3/f_0 \simeq 3$ ms) at high viscosities \cite{singer:EF2018}. Remarkably, the universal scaling $T_{1LM}\propto f_0$ found for all bitumen and polymers in the slow-motion regime (i.e. $\omega_0 \tau_R \gg 1$) is also found for $T_{1S}$ of heptane at low volume fractions ($\phi_{C7} \lesssim $ 50 vol\%) in the polymer matrix. This frequency dependence is in stark contrast to the traditional BPP model where $T_{1LM}\propto f_0^2$ is predicted at high-viscosities \cite{bloembergen:pr1948}. In the case of bitumen and polymers, a phenomenological model was proposed to account for $T_{1LM}\propto f_0$ dispersion at high viscosities \cite{singer:EF2018}. The results in Fig. \ref{fg:T1ST2S}(a) indicate that the same phenomenological model may apply to $T_{1S}$ for heptane under nano-confinement in an organic matrix. 

Fig. \ref{fg:T1ST2S}(b) shows that $T_{2S}$ has much less dispersion, as expected. The increase in $T_{2S}$ from low to high $f_0$ (i.e. from the fast- to slow-motion regime) is given by 10/3, independent of the details in $J_{R,T}(\omega)$. The factor 10/3 can be calculated by comparing Eq. \ref{eq:T2RT} in the fast-motion regime ($\omega_0 \tau \ll 1$) to the slow-motion regime ($\omega_0 \tau \gg 1$) \cite{cowan:book}. The resulting $T_{1S}/T_{2S}$ ratio is shown in Fig. \ref{fg:T1ST2S}(c), where $T_{1S}/T_{2S} \simeq $ 20 at $f_0 = $ 400 MHz for $\phi_{C7} = $ 5 vol\%.

\subsection{Surface relaxivity of heptane}

\begin{figure}[h!]
\centering
\includegraphics[width=1\columnwidth,trim=0 1.78cm 0 0, clip]{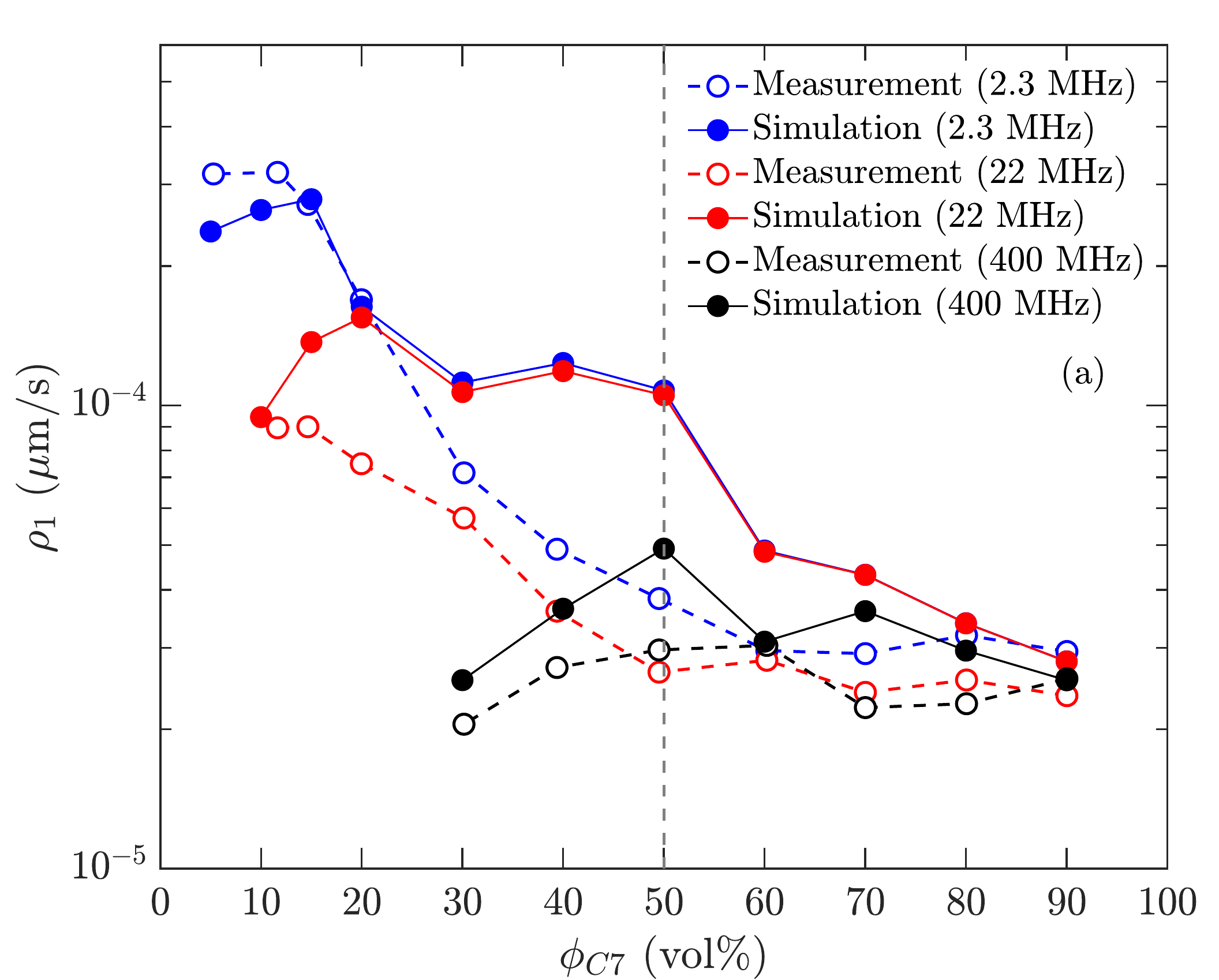} \includegraphics[width=1\columnwidth,trim=0 1.78cm 0 0.3cm, clip]{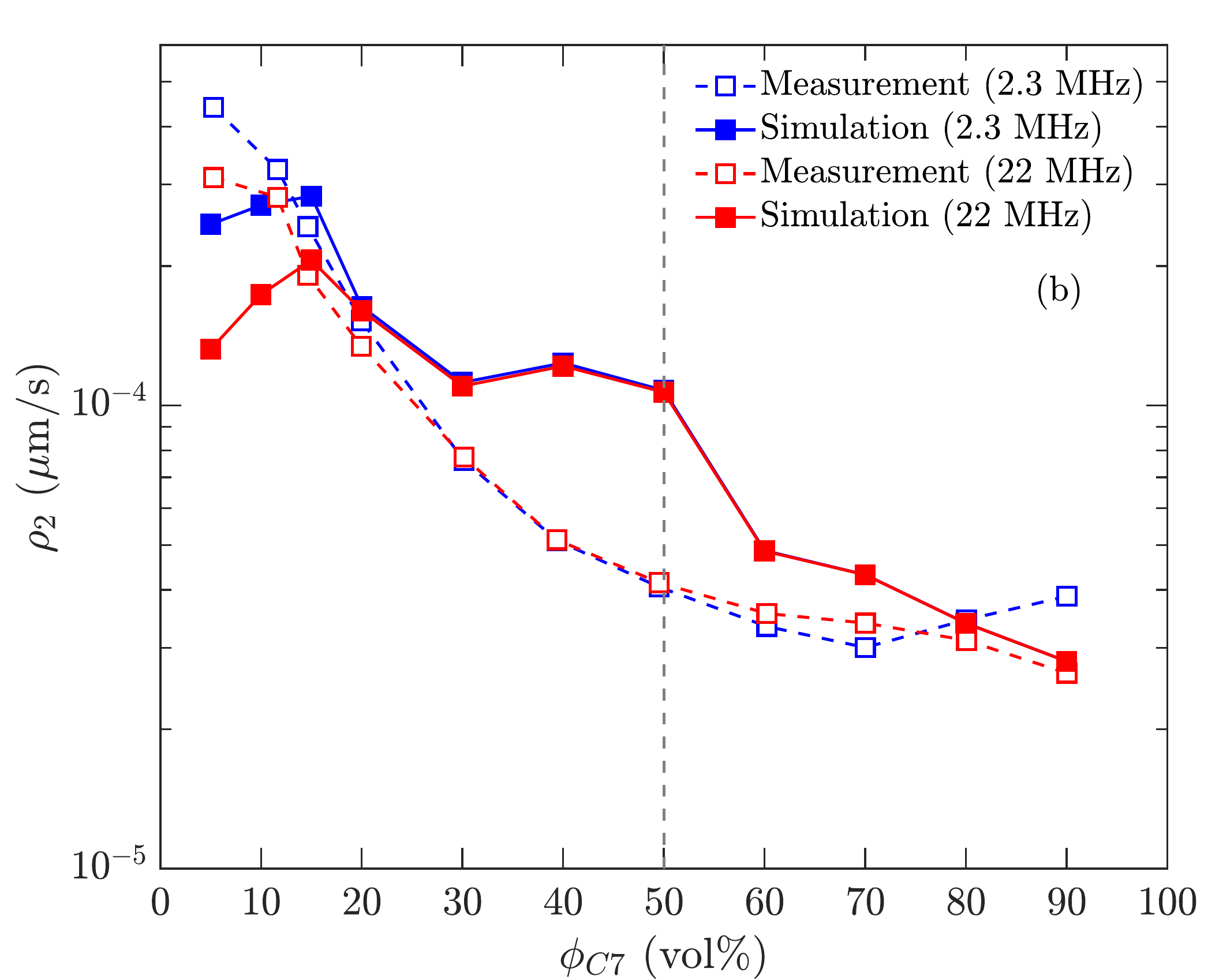} 
\includegraphics[width=1\columnwidth,trim=0 0 0 0.3cm, clip]{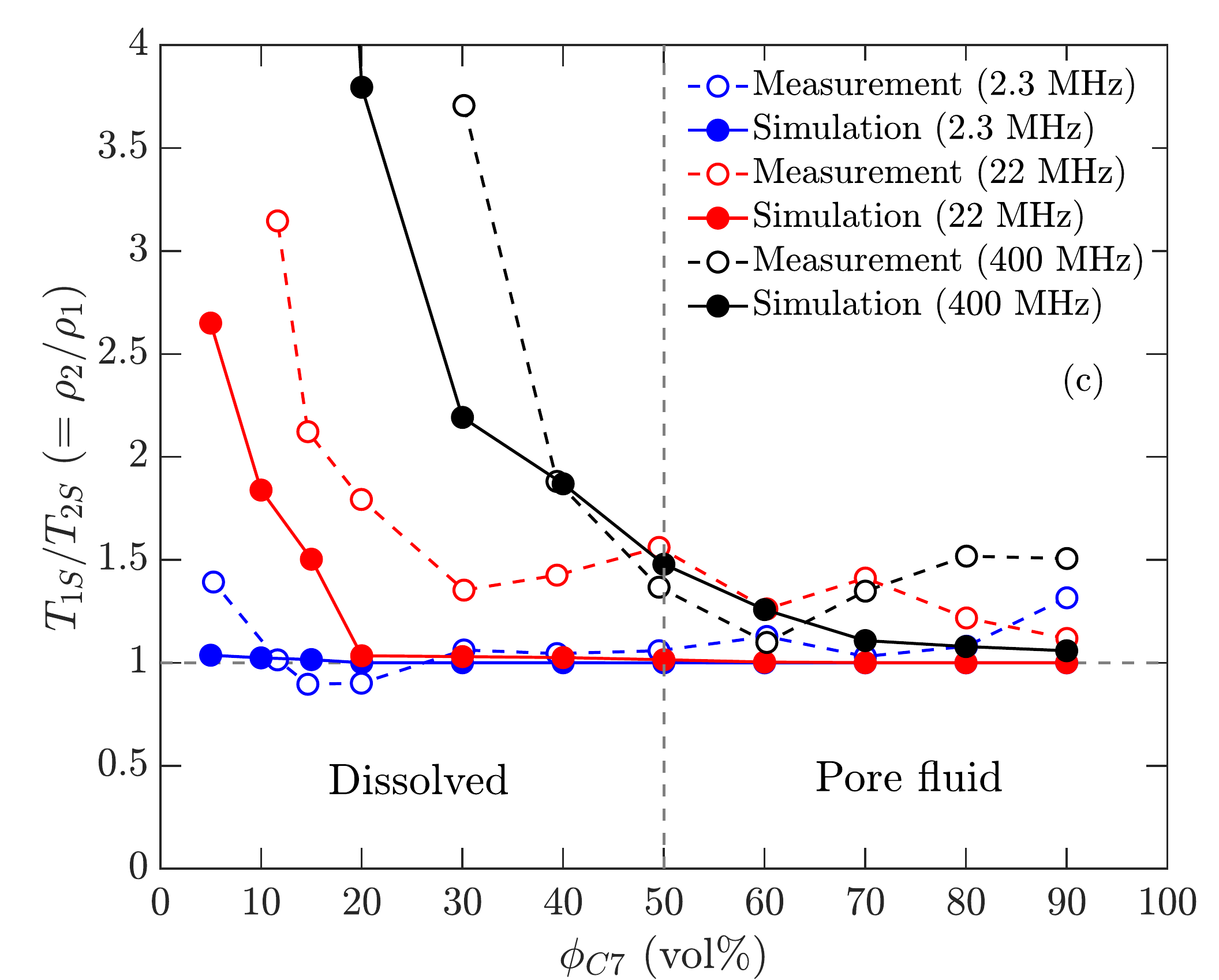}
\caption{Surface relaxivities (a) $\rho_{1}$, (b) $\rho_{2}$, and (c) $T_{1S}/T_{2S}\, (=\rho_2/\rho_1) $ ratio as a function of heptane volume fractions $\phi_{C7}$ for both simulations (closed symbols) and measurements (open symbols), at frequencies $f_0$ = 2.3 MHz, 22 MHz, and 400 MHz. Dashed vertical line shows dissolved heptane region $\phi_{C7} < 50$ vol\%, and pore fluid region $\phi_{C7} > 50$ vol\%.}
\label{fg:Rho}
\end{figure}

The surface-relaxivity parameter $\rho_{1,2}$ is given by the following expression \cite{brownstein:pr1979}:
\begin{eqnarray}
\frac{1}{T_{1S,2S}} &=& {\rho_{1,2}}  \frac{S}{V_p} \, \label{eq:brown}
\end{eqnarray}
where $V_p$ is the pore volume, $S$ is the surface area of the pore, and $\rho_{1,2}$ are the surface-relaxivity parameters. $S$ and $V_p$ incorporate the geometric factors related pore geometry, while $\rho_{1,2}$ incorporate the surface interactions between heptane and the polymer surfaces. The surface to pore-volume ratio of the polymer matrix is related to the surface to grain-volume ratio of the polymer as such:
\begin{eqnarray}
\frac{S}{V_p} &=&  \frac{1-\phi_{C7}}{\phi_{C7}} \frac{S}{V_g}  = \frac{4}{d}. \label{eq:S_Vp}
\end{eqnarray}
$V_g$ is the grain volume of the polymer, which MD simulations have previously shown is $S/V_g \approx 0.859$~${\rm \AA}^{-1}$ for branched alkanes \cite{lekomtsev:rjgc2002}, independent of the chain length. $d$ is the equivalent diameter of a cylindrical pore shown in Fig.~\ref{fg:Porefluid}, where we imagine heptane to be extended in a cylindrical pore at high confinement. The diameter of the extended heptane is around $d = 4.2 {\rm \AA}$, which corresponds to $\phi_{C7}=50$ vol\% according to Eq. \ref{eq:S_Vp}. It is fair to assume that below $\phi_{C7} < 50$ vol\%, heptane molecules interact mainly with the polymer surfaces, and can be thought of as being \textit{ab}sorbed (i.e. dissolved) in the polymer matrix. Note that Eq. \ref{eq:brown} is valid in the small-pore regime, otherwise known as the ``fast-diffusion" regime, where $\rho_{1,2} \, d/D_0 \ll 1$ holds. The fast-diffusion regime holds in the present case, as it also does for nano-pore systems such as those found in shale.

Using Eq. \ref{eq:brown} and \ref{eq:S_Vp} with constant $S/V_g $ results in the following expressions:
\begin{eqnarray}
\frac{1}{\rho_{1,2}} &=& {T_{1S,2S}}  \frac{1-\phi_{C7}}{\phi_{C7}} \frac{S}{V_g} \label{eq:Rho}, \\
\frac{T_{1S}}{T_{2S}} &=& \frac{\rho_{2}}{\rho_{1}}.
\end{eqnarray}
The resulting $\rho_{1}$ and $\rho_{2}$ are plotted in Figs. \ref{fg:Rho}(a) and (b), respectively, while the ratio is plotted in Fig. \ref{fg:Rho}(c). Also shown in Fig. \ref{fg:Rho} is the separation between dissolved and pore-fluid states at $\phi_{C7}=50$ vol\%. The simulations show that the surface-relaxivities $\rho_1$ and $\rho_2$ are independent of $\phi_{C7}$ and $f_0$ for $\phi_{C7} \gtrsim $ 50 vol\%, as expected in conventional pores. However below $\phi_{C7} \lesssim $ 50-60 vol\%, both $\rho_1$ and $\rho_2$ increase with decreasing $\phi_{C7}$, which we interpret as the ``dissolved" region where heptane is no longer in contact with other heptane molecules due to increased confinement in the polymer matrix. We also find that $\rho_1$ decreases with increasing $f_0$, i.e. is dispersive, in the dissolved region. The simulations also show that $T_{1S}/T_{2S} \simeq 1$ in the conventional pore-fluid region, while $T_{1S}/T_{2S} \gtrsim 4 $ in the dissolved region which is consistent with previously reported measurements of light hydrocarbons dissolved in kerogen and bitumen.

Fig. \ref{fg:Rho} also shows measurements of $\rho_1$, $\rho_2$, and $T_{1S}/T_{2S}$, where the trends are consistent with simulations. $T_{1S}/T_{2S}$ is found to be systematically lower for simulations compared to measurements. This discrepancy is attributed to the fact that the simulated polymer in the mix has a lower molecular weight $M_w = $ 912 g/mol ($\eta \simeq $ 1000 cP at ambient temperatures) compared to the measured polymer in the mix $M_w$ = 9436 g/mol ($\eta \simeq  $ 333$\,$400 cP at ambient temperatures). Nevertheless, both simulations and measurements indicate that $T_{1S}/T_{2S}$ increases with increasing confinement and increasing frequency, indicating that $^1$H-$^1$H dipole-dipole relaxation enhanced by nano-pore confinement is the dominant surface-relaxation mechanism in saturated organic-rich shales.

\section{Conclusion}\label{sc:Conclusions}

We report on MD simulations of heptane confined in a polymer-heptane mix as a function of heptane volume fraction $\phi_{C7}$ in the mix. Our motivation for studying this system is that the high-viscosity polymer acts as a model of kerogen and bitumen, where a decrease in $\phi_{C7}$ results in an increase in confinement of heptane in the transient organic ``nano-pores" of the polymer matrix. MD simulations of the restriction in translational diffusion coefficient $D_T/D_0$ of heptane in the polymer-heptane mix indicates a power-law dependence $D_T/D_0 \simeq \phi_{C7}^{m-1}$, with an Archie cementation exponent of $m\simeq 3.68$. The simulations agree well with NMR measurements ($m\simeq 3.44$) on similar systems. Furthermore, these findings are consistent with previously reported measurements of water in immature kerogen isolates, which indicate that the high-viscosity polymer is a good model for immature kerogen.

We then report on MD simulations of $^1$H NMR $T_1$ and $T_2$ from $^1$H-$^1$H dipole-dipole interactions for heptane in a polymer-heptane mix, as a function of heptane volume fraction $\phi_{C7}$ and NMR frequency $f_0$. The simulations naturally separate the contributions from intra-molecular $T_{1R,2R}$ (from rotational diffusion) versus inter-molecular relaxation $T_{1T,2T}$ (from translational diffusion). It is found that intra-molecular relaxation dominates over inter-molecular (i.e. $T_{1T,2T}/T_{1R,2R} > 1$) above $\phi_{C7} \gtrsim $ 70 vol\%. Below $\phi_{C7} \lesssim $ 70 vol\%, inter-molecular relaxation dominates (i.e. $T_{1T,2T}/T_{1R,2R} < 1$), except at high frequencies $f_0 \gtrsim $ 400 MHz where the reverse is found for $T_1$ relaxation (i.e. $T_{1T}/T_{1R} > 1$).

MD simulations of the total relaxation $T_1$ and $T_2$ are found to monotonically decrease with decreasing $\phi_{C7}$ as a result of increasing confinement of heptane in the mix. The MD simulations are found to be consistent with $T_1$ and $T_2$ measurements at $f_0$ = 2.3 MHz, 22 MHz, and 400 MHz. Good agreement is found between measurements and simulation, except in the region around $\phi_{C7} \simeq $ 50 vol\% where measurements overestimate $T_1$ and $T_2$ compared to simulations. We propose that the overestimate from the measurements is a result of a decrease in the concentration of dissolved oxygen at $\phi_{C7} \simeq $ 50 vol\%, which is qualitatively confirmed by independent MD simulations of the solubility of oxygen in the mix.

We use the MD simulation results to compute the surface-relaxation components $T_{1S}$ and $T_{2S}$ of heptane in the polymer ``pores". The simulations show that $T_{1S}$ is dispersive above $f_0 \gtrsim $ 10 MHz, and furthermore that $T_{1S}$ tends towards the functional form $T_{1S}\propto f_0$ (specifically $T_{1S} \times 2.3/f_0 \simeq 7$ ms) for $\phi_{C7} \lesssim $ 50 vol\% and high frequencies $f_0 \gtrsim $ 500 MHz. Remarkably, this functional form of the dispersion is consistent with the previously reported measurements of polymers and bitumen where $T_{1LM}\propto f_0$ (specifically $T_{1LM} \times 2.3/f_0 \simeq 3$ ms) at high viscosities, which is in stark contrast to the traditional BPP model where $T_{1LM}\propto f_0^2$ is predicted at high viscosities \cite{bloembergen:pr1948}. In the case of bitumen and polymers, a phenomenological model was proposed to account for $T_{1LM}\propto f_0$ dispersion at high viscosities \cite{singer:EF2018}. Our findings suggest that the same phenomenological model may apply to the surface relaxation of light hydrocarbons in an organic nano-confined matrix.

The simulations show that the surface-relaxivities $\rho_1$ and $\rho_2$ are independent of $\phi_{C7}$ and $f_0$ for $\phi_{C7} \gtrsim $ 50 vol\%, as expected in conventional pores. However below $\phi_{C7} \lesssim $ 50-60 vol\%, both $\rho_1$ and $\rho_2$ increase with decreasing $\phi_{C7}$, which we interpret as the ``dissolved" region where heptane is no longer in contact with other heptane molecules due to increased confinement in the polymer matrix. We also find that $\rho_1$ decreases with increasing $f_0$, i.e. it is dispersive, in the dissolved region. The simulations also show that $T_{1S}/T_{2S} \simeq 1$ in the conventional pore-fluid region, while $T_{1S}/T_{2S} \gtrsim 4 $ in the dissolved region which is consistent with previously reported measurements of light hydrocarbons dissolved in kerogen and bitumen. 

Measurements of $\rho_1$, $\rho_2$, and $T_{1S}/T_{2S}$ show consistent trends with simulations as a function of $\phi_{C7}$ and $f_0$, however $T_{1S}/T_{2S}$ is found to be systematically lower for simulations compared to measurements. This discrepancy is attributed to the fact that the simulated polymer in the mix has a lower molecular weight compared to the measured polymer in the mix. Nevertheless, both simulations and measurements indicate that $T_{1S}/T_{2S}$ increases with increasing confinement and increasing frequency, indicating that $^1$H-$^1$H dipole-dipole relaxation enhanced by nano-pore confinement is the dominant surface-relaxation mechanism in saturated organic-rich shales.

\section{Acknowledgments}

We thank Chevron, the Rice University Consortium on Processes in Porous Media, and the American Chemical Society Petroleum Research Fund (No. ACS-PRF- 58859-ND6) for funding this work. We gratefully acknowledge the National Energy Research Scientific Computing Center, which is supported by the Office of Science of the U.S. Department of Energy (No. DE-AC02-05CH11231), and the Texas Advanced Computing Center (TACC) at The University of Texas at Austin, for HPC time and support.

\appendix
\section{Effect of dissolved oxygen on measurements}\label{SI:O2conc}

Solubility is determined by the excess chemical potential. Here we predict the excess chemical potential of O$_2$ in the alkane/polymer mixture using 
\begin{eqnarray}
\beta \mu^{\rm ex}_{\mathrm{O_2}} = \ln x_0 - \ln p_0 - \ln \langle e^{-\beta \Delta U}\rangle_0 \, 
\end{eqnarray}
which is the quasichemical organization of the potential distribution theorem \cite{paulaitis2002hydration,beck2006potential,Pratt2004Free}. In the above equation, $\ln x_0$ is the work required to move the solvent out of the inner shell defined around the oxygen molecule, $-\ln p_0$ is the work required to created the empty inner shell in the solvent (in the absence of the solute), and $- \ln \langle e^{-\beta \Delta U}\rangle_0$ is the contribution from the interaction of the oxygen with the rest of the solvent when the inner shell is empty. Exploratory calculations show that the interaction between O$_2$ and alkane/polymer matrix is dominated by inner-shell exclusion (steric effects) and long-range van~der~Waals interactions. To this end, we choose an inner shell cavity that is large enough to accommodate the solute but small enough such that $\ln x_0 = 0$. We make a conservative choice of 2.9~{{\rm \AA}} for the inner shell radius. Here O$_2$ was modeled using the three site model that has both partial charge and dispersion contribution {\bf Ref}. 

For computing $p_0$, we first define a cubic grid of size $13\times 13\times 13$~{{\rm \AA}}$^3$. The grid sites are separated by 3~{{\rm \AA}}. (The simulation boxes are all about 40~{{\rm \AA}} and hence the grid sits entirely within the simulation cell.) Using the grid sites as reference, we find the number of occurrences for which no carbon atom of the solvent is within 2.9~{{\rm \AA}} of the grid site. All such sites are archived for further analysis. This calculation also directly provides the probability $p_0$ of finding a cavity of size 2.9~{{\rm \AA}} in the hydrocarbon matrix. For the cavities archived from the study above, we compute $- \ln \langle e^{-\beta \Delta U}\rangle_0$ by particle insertion \cite{paulaitis2002hydration,beck2006potential,Pratt2004Free,Widom1982}. For these calculations, based on the convergence of the free energy, we used only a smaller subset (up to 1000 frames) of the overall 5000 frames. Note that for each site, we also consider three random orientations of the oxygen molecule, further enhancing the statistical reliability. 

\begin{figure}[ht!]
\centering
\includegraphics[width=1\columnwidth]{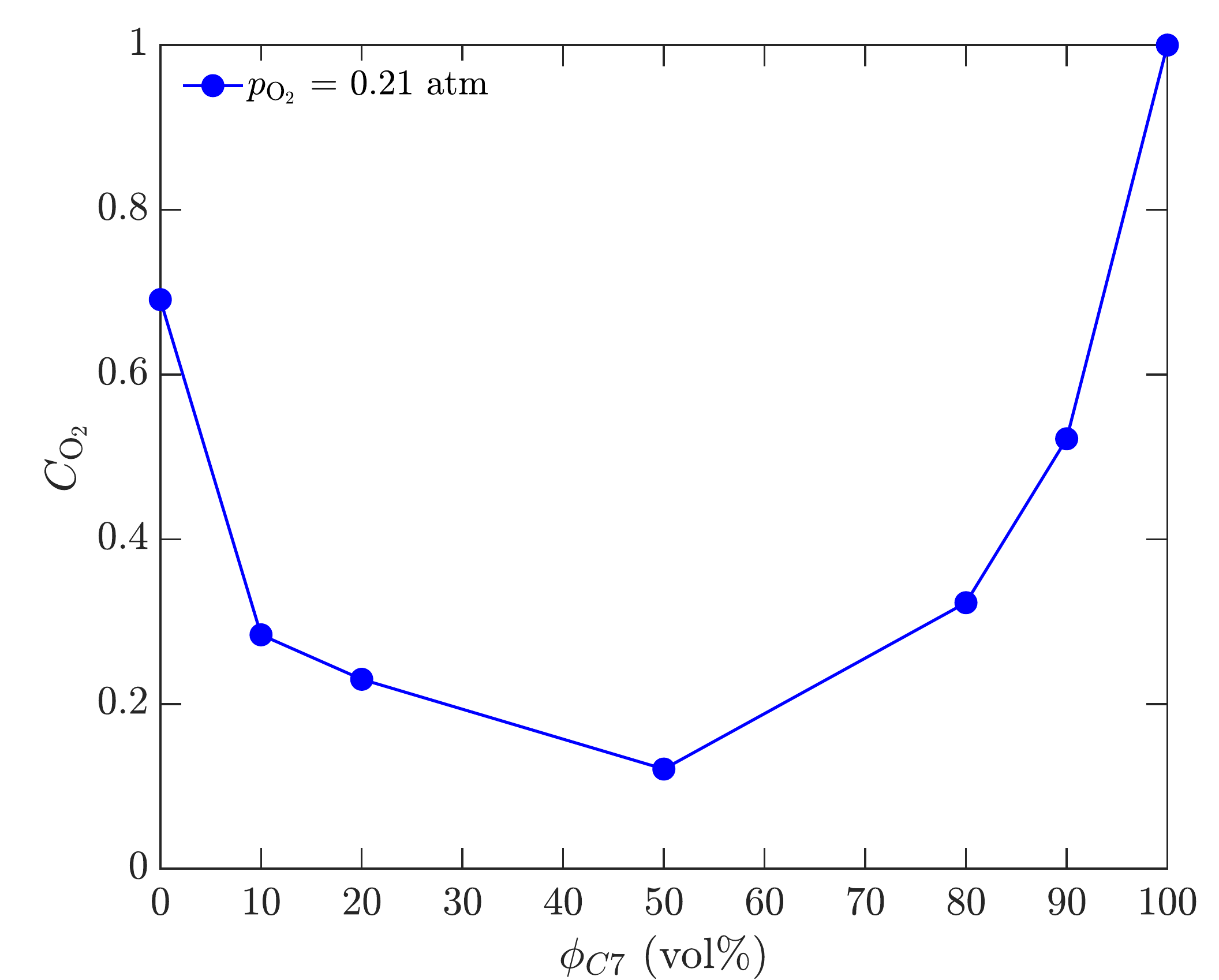} 
\caption{MD simulations of concentration $C_{\rm O_2}$ of dissolved oxygen in the polymer-heptane mix, as a function of heptane volume fraction $\phi_{C7}$. $C_{\rm O_2}$ is defined relative to pure heptane ($\phi_{C7} = 100$ vol\%), under ambient conditions.} \label{fg:O2}
\end{figure}

We obtain the solubility of oxygen in heptane to be 676 ppm. The experiments suggest that the solubility of oxygen in $n$-heptane is around 132 ppm (by weight) \cite{NIST_O2_hept,hesse1996o2sol}. To obtain this value (assuming oxygen partial pressure of 0.21~atm.), we need $\beta\mu^{\rm ex}_{\mathrm{O_2}} \approx 5.0$. In energy units, the difference between our computed value and 5.0 is about $0.9$~kcal/mol. This small difference may result from deficiencies of the force field itself. However, we suspect the relative solubility trends to be well-captured by our simulations. Since the solubility relative to the bulk is of most interest, in Fig.~\ref{fg:O2} we show the relative solubility of O$_2$ in the alkane/polymer mixture. Considering the relative solubility also serves to minimize the errors in the absolute solvation values, that are off by a factor of 5.

The relation between the measured $T_{1,2}^{\rm meas}$ and the intrinsic $T_{1,2}$ of interest is given by the following expression \cite{teng:jmr2001}:
\begin{align}
\frac{1}{T^{\rm meas}_{1,2}}  &= \frac{1}{T_{1,2}}  + \frac{C_{\rm O_2}}{T_{1{\rm O_2},2{\rm O_2}}}. \label{eq:T1meas}
\end{align}
The measured $T_{1{\rm O_2}} \, (= T_{2{\rm O_2}} )$ for pure heptane at ambient conditions are $T_{1{\rm O_2}} = $ 2490 ms, 2620 ms, and 5580 ms at $f_0 = $ 2.3 MHz, 22 MHz, and 400 MHz, respectively \cite{singer:EF2018}. It was previously shown that $T_{1{\rm O_2}}$ is roughly constant for solvents with the molecular weight of heptane or higher \cite{teng:jmr2001}. In other words, $T_{1{\rm O_2}} \, (= T_{2{\rm O_2}} )$ in Eq. \ref{eq:T1meas} is assumed to be independent of $\phi_{C7}$. As shown in Fig. \ref{fg:Crossplot}, we find good agreement between simulated $T_{1,2}$ and measured $T_{1,2}$ assuming $C_{\rm O_2}$ = 1 for all $\phi_{C7}$ in Eq. \ref{eq:T1meas}. 

\begin{figure}[ht!]
\centering
\includegraphics[width=1\columnwidth]{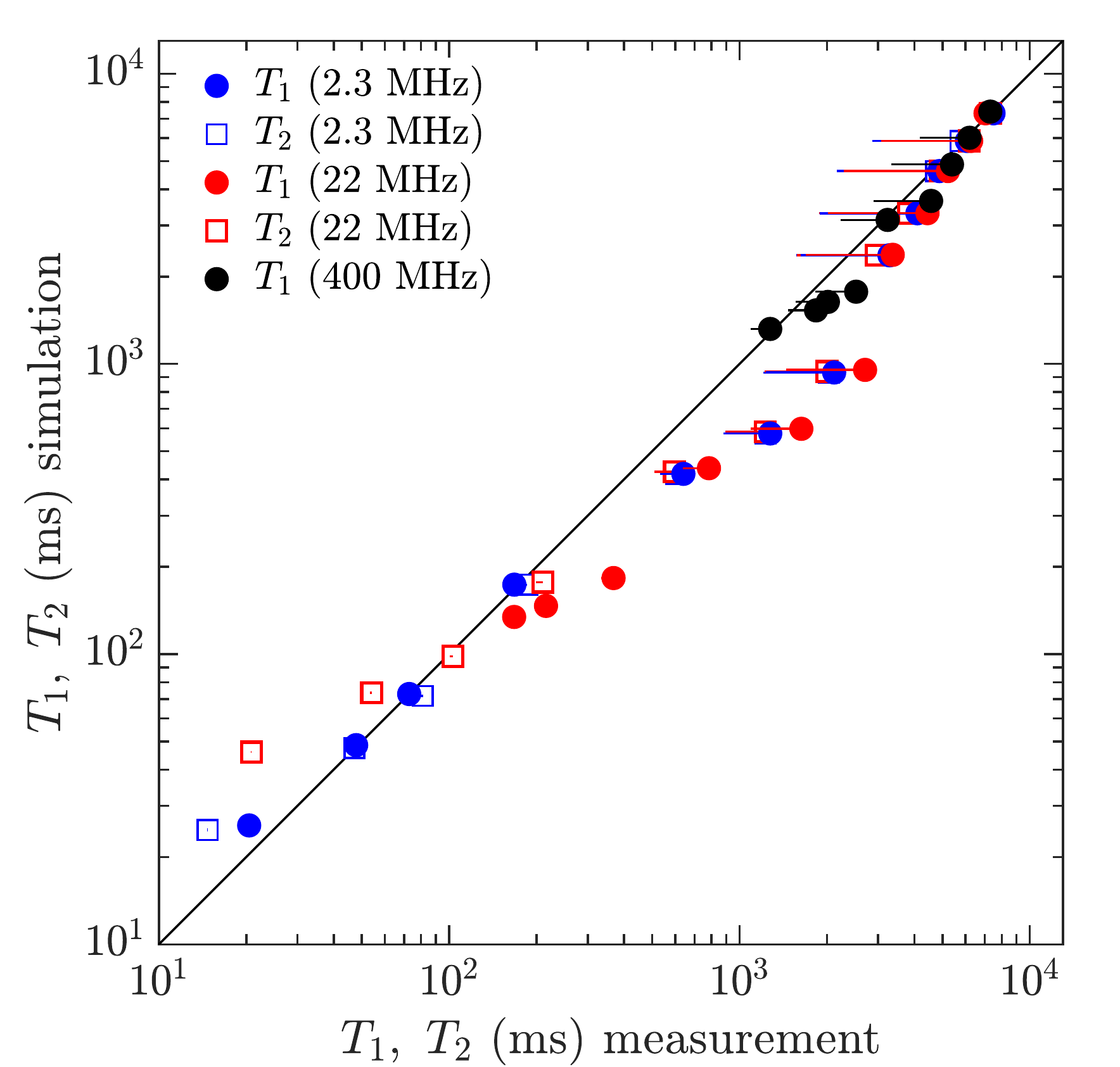} 
\caption{Correlation cross-plot of measurements vs. simulations of $T_1$ and $T_2$ at $f_0$ = 2.3 MHz, 22 MHz, 400 MHz, for various heptane volume fractions $\phi_{C7}$. Symbols use $C_{\rm O_2}$ = 1 in Eq. \ref{eq:T1meas} to determine $T_{1,2}$ from measured $T_{1,2}^{\rm meas}$, while leftmost point of the individual vertical lines uses $C_{\rm O_2}$ values from Fig. \ref{fg:O2}.}
\label{fg:CrossplotO2}
\end{figure}
\textbf{}

However, as shown in Fig. \ref{fg:O2}, $C_{\rm O_2}$ decreases at around $\phi_{C7} \simeq $ 50 vol\%, and therefore the assumption that $C_{\rm O_2} = 1$ for all $\phi_{C7}$ may not be accurate. In order to quantify this effect, Fig. \ref{fg:CrossplotO2} shows the measured $T_{1,2}$ using the simulated $C_{\rm O_2}$ values as a function of $\phi_{C7}$ from Fig. \ref{fg:O2}, the results of which are shown as the leftmost point of the vertical lines in Fig. \ref{fg:CrossplotO2}. Using the $C_{\rm O_2}$ values from Fig. \ref{fg:O2} improves the comparison between measurements and simulations in the region $\phi_{C7} \simeq $ 50 vol\%, however a discrepancy is then found in the region $\phi_{C7} > $ 70 vol\%. We note however that the simulated $C_{\rm O_2}$ in Fig. \ref{fg:O2} are qualitative and designed to capture the overall trends in $C_{\rm O_2}$, namely that $C_{\rm O_2}$ decreases around $\phi_{C7} \simeq $ 50 vol\%, which coincides with the discrepancy between measurements and simulations of $T_{1,2}$ in that region. This gives credibility, though not certainty, to the proposition that variations in $C_{\rm O_2}$ with $\phi_{C7}$ are the cause of the discrepancy between measurements and simulations at $\phi_{C7} \simeq $ 50 vol\%. 

\clearpage
\section{Diffusivity of Water in Kerogen}\label{SI:DT2expt}
\begin{figure}[ht!]
\centering
\includegraphics[width=1\columnwidth]{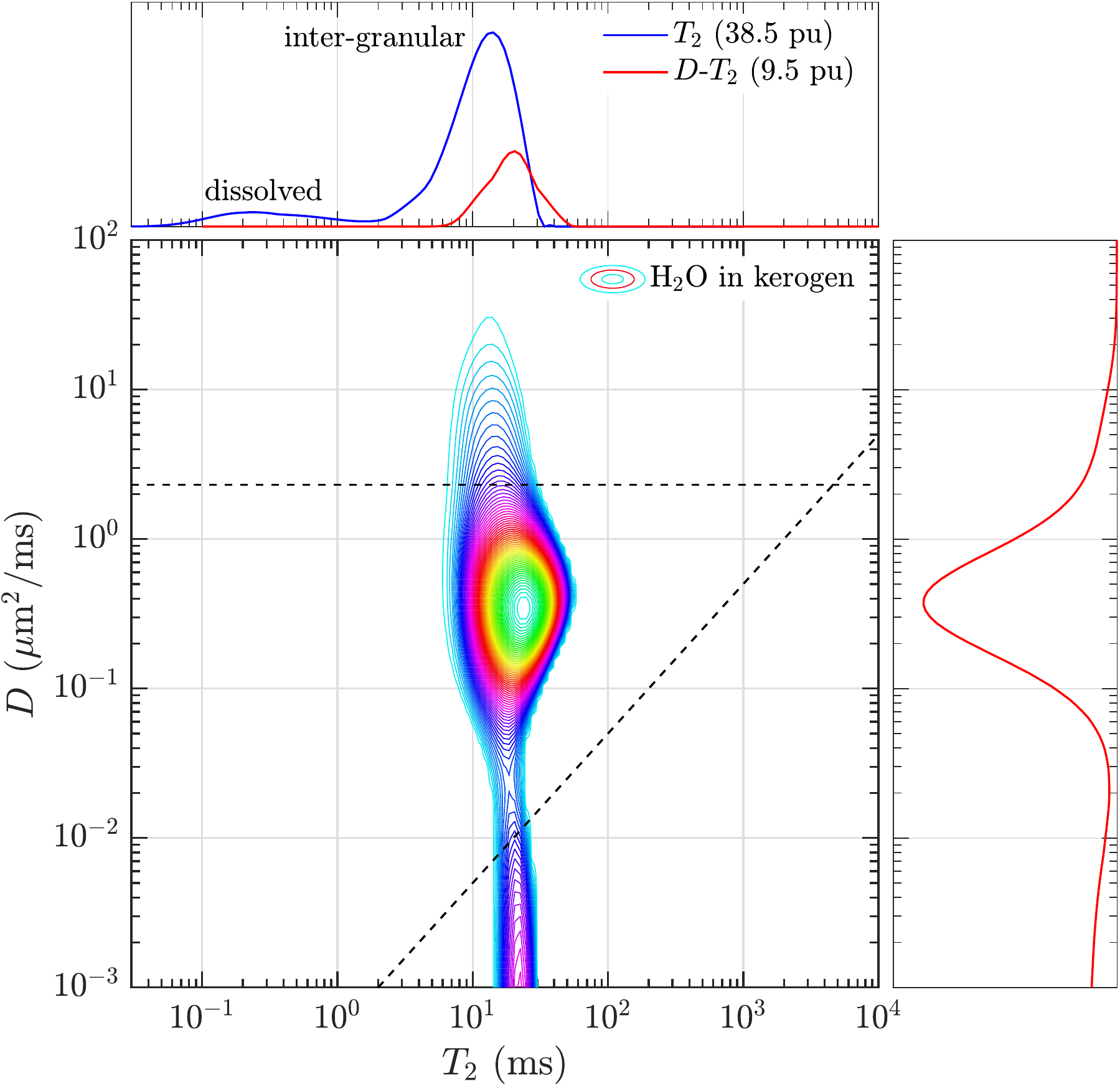} 
\caption{Diffusion-$T_2$ ($D$-$T_2$) measurement at ambient of water-saturated isolated kerogen pellets from a Kimmeridge outcrop (same kerogen as used in \cite{singer:petro2016,chen:petro2017}). Right panel shows $D$ projection (red). Upper panel shows the $T_2$ projection from $D$-$T_2$ (red), along with full $T_2$ distribution (blue). The projection from $D$-$T_2$ (9.5 pu) shows less signal intensity than the full $T_2$ (35.8 pu) due to limitations in the $D$-$T_2$ measurement. The diffusion coefficient (taken at the peak of the $D$ distribution) of the inter-granular water is detectable, while the diffusion coefficient for dissolved water is not detectable. Dashed horizontal line is the bulk $D_0$ for water, while the dashed diagonal line is the bulk alkane line \cite{lo:SPE2002}.}
\label{fg:DT2}
\end{figure}
\clearpage


\end{document}